\documentclass[%
reprint,
 amsmath,amssymb,
 aps,
floatfix,
]{revtex4-1}

\usepackage{graphicx}
\usepackage{dcolumn}
\usepackage{amsmath}
\usepackage{bm}
\usepackage{color} 
\usepackage{braket}
\usepackage[normalem]{ulem}
\usepackage{gensymb}
\usepackage[dvipsnames]{xcolor}
\usepackage{lpic}
\usepackage{MnSymbol}
\usepackage{blindtext}

\usepackage{tikz}
\usepackage[percent]{overpic}
\usepackage{stackengine}
\usepackage{physics}
\usepackage{siunitx}
\usepackage{textcomp}
\usepackage{slashbox}

\begin{document}

\preprint{APS/123-QED}

\title{A Voigt effect based 3D vector magnetometer}

\author{Tadas Pyragius$^1$}
\author{Hans Marin Florez$^{1,2}$}
\author{Thomas Fernholz$^{1}$}
\email{thomas.fernholz@nottingham.ac.uk}
\affiliation{
$^{1}$School of Physics \& Astronomy, University of Nottingham, University Park, Nottingham NG7 2RD, UK\\
$^{2}$Instituto de F\'{\i}sica, Universidade de S\~ao Paulo, 05315-970 S\~ao Paulo, SP-Brazil
}

\date{\today}

\begin{abstract}
We describe a method to dispersively detect all three vector components of an external magnetic field using alkali atoms based on the Voigt effect. Our method relies on measuring the linear birefringence of the radio frequency dressed atomic medium via polarization homodyning. This gives rise to modulated polarization signals at the first and second harmonic of the dressing frequency. The vector components of the external magnetic field are mapped onto the quadratures of these harmonics. We find that our scheme can be utilised in both cold and hot atomic gases to detect such external fields in shielded and unshielded environments. In the shielded hot vapour case we achieve field sensitivities in the pT$/\sqrt{\textrm{Hz}}$ range for all 3 vector components, using pump-probe cycles with 125 Hz repetition rate, and limited by the short coherence time of the cell. Finally, our scheme has a simple single axis beam geometry making it advantageous for miniature magnetic field sensors. 
\end{abstract}

\pacs{Valid PACS appear here}
\maketitle

\section{\label{sec:intro}Introduction}

Optically pumped magnetometers, (OPMs), have increasingly been in the spotlight for their broad span of applications ranging from fundamental physics experiments to medical physics. Examples include measurements of the electric dipole moment (EDM)~\cite{edm1,edm2} and searches for exotic physics~\cite{cptviolation} as well as magneto-encephalography (MEG)~\cite{meg1,meg2} and magneto-cardiography~\cite{mcg0,mcg1,mcg2} where detection of the small magnetic fields of the brain and the heart is required. A review can be found in ~\cite{Budker2007}.
In recent years OPMs have become the state-of-the-art magnetic field sensors achieving sub fT$/\sqrt{\textrm{Hz}}$ sensitivity and surpassing the well established SQUID based sensors~\cite{Romalis2002,Romalis2010,Romalis2013}. 
In its simplest operation, an OPM uses a pump-probe laser to measure the atomic Larmor frequency, i.e.\ the frequency of spin precession, by interacting with optically pumped atoms, which in effect measures the strength of the external magnetic field. 
However, in a larger range of applications, a complete determination of the magnetic field is required. Some schemes employ a scalar magnetometer to run as a vector magnetometer by applying a rotating low frequency bias magnetic field~\cite{Yakobson2004,Gao2016}. Another possible approach uses multiple radio-frequency modulations to map the three vector components onto the harmonics of the signal~\cite{Romalis2004,Budker2014}. The effects of the field orientation on the resulting signal phase have been studied for different configurations of a modulating field and may be used for full vector magnetometry~\cite{ingleby}.
Also an all-optical scheme with crossed beams was demonstrated to extract the three field components~\cite{Gao2015}. 
To date, most of the OPM schemes are based on pump-probe configurations that rely on the Faraday rotation, i.e.\ circular birefringence of the medium. As a result, the majority of such schemes require an orthogonal pump-probe geometry for high efficiency of detection \cite{Romalis2002}. However, this geometry is not convenient for developing miniature sensors, whilst a parallel configuration is compatible with chip-scale and compact atomic magnetometers. The issue can be overcome by introducing modulation fields, but several fields are required to extract the full information.

In this paper we demonstrate an alternative 3D vector magnetometer based on the measurement of the Voigt effect, i.e.\ linear birefringence arising from aligned rather than oriented spin states. It is measured with a probe beam detuned from an optical resonance, in the presence of a single radio-frequency field~\cite{Jammi}. Previous work on double resonance detection of aligned states measured linear dichroism on a resonant optical transition \cite{Weis1,Weis2}.
The method presented here, maps the three vector components of the external field, detected via demodulation of the probe beam's ellipticity, onto orthogonal quadratures of the first and second harmonic of the dressing frequency. State preparation and detection are performed in a parallel pump-probe geometry. We demonstrated vector capability of our magnetometer over a range of $\pm0.3$~nT longitudinal and $\pm180$~nT transverse fields and analyzed its sensitivity in a shielded environment in open loop operation. Active feedback on the external field should enable an extension of dynamic range as well as operation in unshielded scenarios.

The paper is organized as follows. In Section~\ref{sec:theory}, we briefly describe the linear birefringence induced by radio-frequency dressed states. This provides predictions for the mapping of field components onto orthogonal quadratures of the first and second harmonics of the rf oscillation in the signal's response. In the following part of the paper, we report on the experimental realization using two different types of atomic ensemble. Section~\ref{sec:expColdatoms} demonstrates the detection principle with laser cooled atoms, prepared in a pure quantum state. Section \ref{sec:expvapour} describes the extension to a magnetically shielded vapour cell by combining the Voigt effect with synchronous pumping. Experimental results on vector sensitivity are shown together with an analysis of noise performance. Section~\ref{sec:Conclusions} presents our conclusions.

\section{\label{sec:theory}Magnetometry with radio-frequency dressed states}

Optically pumped magnetometers utilise dispersive coupling of light to an atomic ensemble in the presence of an external magnetic field. The Larmor precession of spin-polarized atoms causes a modulation of the medium's birefringence, which can be observed polarimetrically. In our scheme, we actively drive such precession with an additional radio-frequency field. Here, we present a brief description of the driven medium and its interaction with the light field in terms of dressed states, as discussed in our previous work~\cite{Jammi}. Our model includes the dependence on field orientation, which allows for the extraction of full vector information from measurements of either linear or circular birefringence.

We consider atoms interacting with a static field $\mathbf{B}_{\mathrm{dc}}=B_{\mathrm{dc}}\mathbf{e}_z$ and a field oscillating at a radio-frequency $\omega$ in a transverse direction $\mathbf{B}_{\mathrm{rf}}(t)=B_{\mathrm{rf}}\cos(\omega t) \mathbf{e}_x$. For weak fields, the time-dependent interaction Hamiltonian of an atom with spin $\mathbf{F}$ of constant magnitude can be approximated by $\hat{H}=(\mu_B g_F/\hbar) \hat{\mathbf{F}}\cdot\left(\mathbf{B}_{\mathrm{rf}}(\omega t)+\mathbf{B}_{\mathrm{dc}}\right)$,
where $\mu_B$ is the Bohr magneton, $g_F$ is the Land\'e factor, and $\hbar$ is the reduced Planck constant.
Depending on the sign of the $g_F$ factor, 
using positive $\omega$, we transform the Hamiltonian to a frame rotating about the $z$-axis, according to $\hat{H}_\mathrm{rot}=\hat{U}\hat{H}\hat{U}^{-1}+i\hbar^{-1}\left(\partial_t\hat{U}\right)\hat{U}^{-1}$ with a time-dependent rotation operator $\hat{U}=e^{\mathrm{sgn}(g_F) i\omega t \hat{F}_z/\hbar}$.
Neglecting counter-rotating terms, the transformed, effective Hamiltonian takes the form
\begin{equation}
\hat{H}_{\mathrm{eff}}=\frac{\mu_Bg_F}{\hbar}\hat{\mathbf{F}}\cdot\mathbf{B}_{\mathrm{eff}}.\label{eq:Heff}
\end{equation}

\begin{figure}[h!]\hspace*{-0.4cm}
\begin{overpic}[scale=0.3]{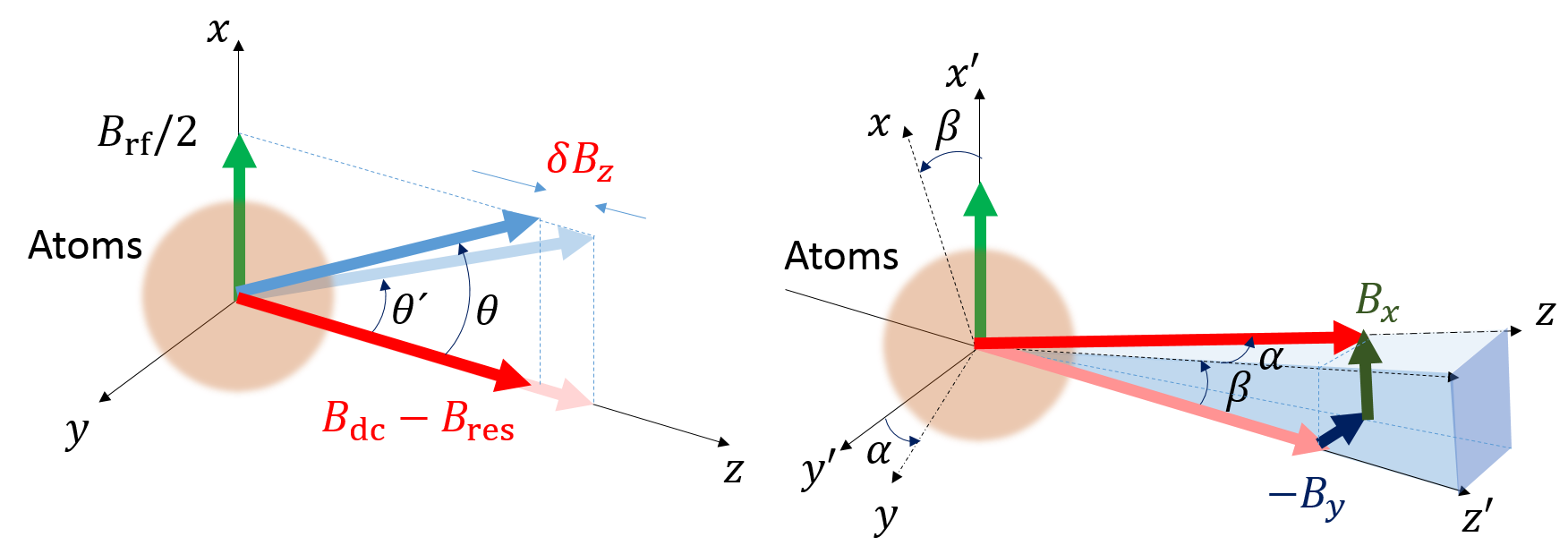}
 \put (3,33) {a)} \put (45,33) {b)}
\end{overpic}    
\caption{Geometrical depiction of the effective field in the rotating frame. (a) The effective field encloses an angle $\theta$ with the $z$-axis. An external field variation $\delta B_z$ changes the angle $\theta \rightarrow \theta'$. (b) The presence of transverse external fields $B_y$ and $B_x$ also changes the orientation of the effective field, rotating it by angles $\alpha$ and $\beta$, respectively.}\label{fig:FrameRotation}
\end{figure}

The effective magnetic field in this frame is given by $\mathbf{B}_{\mathrm{eff}}=B_{\rho}\mathbf{e}_x + (B_{\mathrm{dc}}-B_{\mathrm{res}}) \mathbf{e}_z$, where $B_{\rho}=B_\mathrm{rf}/2$, and $B_{\mathrm{res}}=\pm\hbar\omega/\mu_B g_F$ corresponds to a fictitious magnetic field that defines a resonance condition for the Larmor precession~\cite{Jammi}. As depicted in Fig.~\ref{fig:FrameRotation}(a), the angle enclosed by the effective field and the $z$-axis is 
\begin{equation}
\theta=\frac{\pi}{2}-\mathrm{tan}^{-1}{\frac{B_\mathrm{dc}-B_{\mathrm{res}}}{B_{\rho}}}.
\label{eq:theta}
\end{equation}
E.g., at resonance, i.e.\ for $\theta=\pi/2$, the effective field is orthogonal to the static field $\mathbf{B}_\mathrm{dc}$, pointing in the rotating frame's $x$-direction. 

The eigenstates of the effective, rotating-frame Hamiltonian, i.e.\ the dressed states, can be written as $ \left|\Psi_{\mathrm{rot}}\right\rangle=e^{i\theta\hat{F}_y/\hbar}\left|F,F_z\right\rangle$. But in the laboratory frame the same states are time-dependent, given by $\left|\Psi(t)\right\rangle=\hat{U}^{-1}\left|\Psi_{\mathrm{rot}}\right\rangle$. The dressed states can be prepared directly by synchronous optical pumping \cite{azunish}, or by adiabatic dressing of bare states $\left|F,F_z\right\rangle$, which acquire only a time-dependent phase under transformation to the rotating frame, leading to a different (quasi-) energy. 
In our scheme, magnetometer operation does not rely on a purely dynamical spin evolution, which would be observed as a change in precession frequency. Instead, we assume adiabatic following of dressed states that remain aligned with the effective field. Magnetometry is then enabled by the orientational dependence of the effective field on an additional external field. 

An external field in the $z$-direction enters the spin evolution through the dependence of $\theta$ on the static field strength, see Eq.~(\ref{eq:theta}) and Fig.~\ref{fig:FrameRotation}(a), where we can control the sensitivity $\partial\theta/\partial B_z$ by applying an offset field, such that $B_z=B_\mathrm{offs}+\delta B_z$.
The presence of transverse static fields can also be represented by rotations, as shown in Fig.~\ref{fig:FrameRotation}(b). Field components $B_{y,x}$ rotate the static field about the $x,y$-axes by angles $\alpha$ and $\beta$, respectively. Hence, using a sequential rotation $\mathbf{M}(\alpha,\beta)=\mathbf{R}_x(\alpha)\mathbf{R}_{y}(\beta)$ \cite{Rodrigues}, the atomic spin operator in the laboratory frame is given by $\hat{\mathbf{F}}'=\mathbf{M}(\alpha,\beta)\hat{\mathbf{F}}$, where unprimed coordinates are aligned with the actual field.   
Figure~\ref{fig:FrameRotation} shows that the angles $\alpha$ and $\beta$ are given by
\begin{align}
\alpha&=\arctan\left(\frac{-B_y}{B_z}\cos(\beta)\right), \\
\beta&=\arctan\left(\frac{B_x}{B_z}\right),
\end{align}
with the small angle approximation
\begin{align}
\alpha&\approx\frac{-B_y}{B_z},\  \beta\approx\frac{B_x}{B_z}.
\label{eq:angles}
\end{align}
For a complete description at larger angles, we need to include that the transverse fields increase the actual static field strength to $B_\mathrm{dc}=\sqrt{B_z^2+B_x^2+B_y^2}$, and that the applied rf field is not co-rotated, leading to a reduction of its effective amplitude in the rotating frame, given by $B_\rho=(B_\mathrm{rf}/2)\mathrm{cos}\beta$. 

For the detection of the spin evolution we employ $+45^{\circ}$-linearly polarized probe light propagating in the $z$-direction with a corresponding Stokes parameter $S_y=(c/2)\langle \hat{a}^{\dagger}_{x}\hat{a}_{y}+\hat{a}^{\dagger}_{y}\hat{a}_{x}\rangle$, which is equal to half the photon flux \cite{Jammi}.
The dispersive interaction of the atomic medium with off-resonant light may lead to both circular and linear birefringence, depending on the atomic spin-dependent polarizability tensor. 
After propagation through the medium, neglecting absorption and assuming sufficiently small phase angles, the resulting Faraday and Voigt rotation can be described by Stokes operators 
\begin{align}
\Braket{\hat{S}'_x(t)}&=-G_{F}^{(1)} S_y n_F \Braket{\hat{F}_z(t)},\label{eq:Faraday}\\
\Braket{\hat{S}'_z(t)}&=G_{F}^{(2)} S_y n_F \Braket{\hat{F}_x^2(t)-\hat{F}_y^2(t)}\label{eq:Voigt},
\end{align}
where $\hat{S}'_z$ and $\hat{S}'_x$ 
represent the polarization's rotation and ellipticity as photon flux imbalances of the output light, measured in either a circular or linear basis. 
The coupling strengths $G_{F}^{(k)}$
depend on light detuning, interaction cross section, and the rank-k components of the polarizability tensor. In these equations, we assume interaction with $n_F$ atoms in the same spin state within one hyperfine $F$-manifold and neglect dispersive back-action on the atoms (Stark shifts)~\cite{Jammi}. 

For eigenstates of the effective Hamiltonian, using the geometrical rotations by angles $\alpha$, $\beta$, and $\theta$, we can determine the temporal atomic response in the laboratory frame, measured via Faraday or Voigt effect, in the presence of external magnetic fields.
An adiabatic eigenstate $\left|\Psi_{\mathrm{rot}}\right\rangle$, transformed to the laboratory frame and rotated by $\mathbf{M}(\alpha,\beta)$, leads to spectral decompositions of the measured signals. 

\begin{figure}[t!]
\begin{overpic}[width=0.45\textwidth]{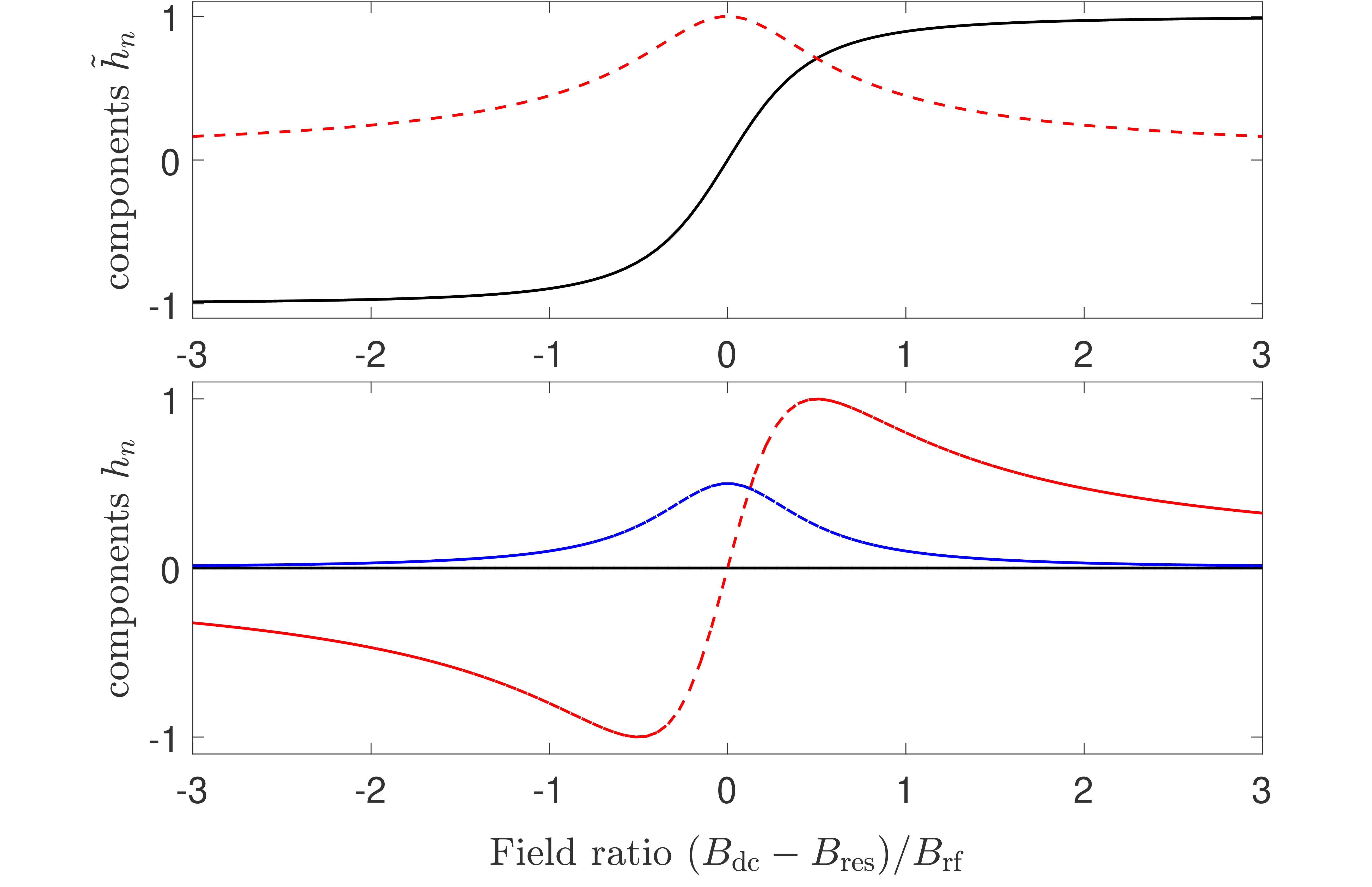}
 \put (0,63) {a)} 
 \put (18,61) {Faraday}
 \put (25,44) {n=0}
 \put (25,52) {\color{red}n=1}
 \put (0,35) {b)}
 \put (18,33) {Voigt}
 \put (59,20) {n=0}
 \put (69,33) {\color{red}n=1}
 \put (59,28) {\color{blue}n=2}
\end{overpic}    
\caption{Spectral decomposition of (a) Faraday rotation proportional to $\Braket{F_z}$ and (b)
Voigt rotation proportional to $\Braket{F_x^2-F_y^2}$, with harmonics $n=0$ (solid black lines), $n=1$ (dashed red lines) and $n=2$ (dashed-dotted blue lines).}
\label{fig:hcomps}
\end{figure}
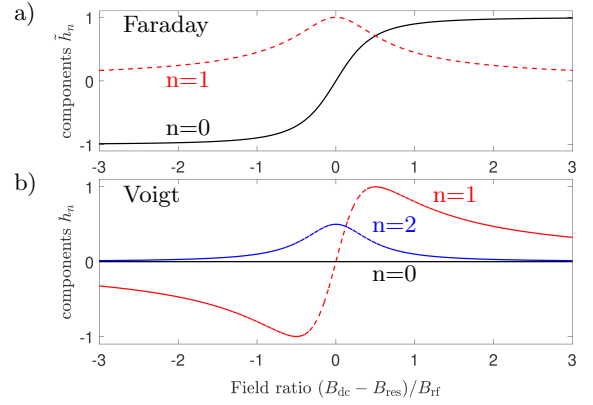

For Faraday rotation, this is given by
\begin{equation}
\Braket{\hat{S'}_x(t)}=-\frac{1}{2}G_F^{(1)}S_y n_F\hbar F_z\sum_{n=0}^1 \tilde{h}_n(\theta)e^{in\omega t}+c.c.,
\label{eqn:spectraloutputFz}
\end{equation}
using $\Bra{F,F_z}\hat{F}_z\Ket{F,F_z}=\hbar F_z$. From Eqs.~(\ref{eq:h0Fz}) and (\ref{eq:h1Fz}), we find the spectral components in the small
angle approximation \begin{align}
    \label{eqn:hcompFzLowangle}
    &\left(
    \tilde{h}_0,\tilde{h}_1\right)^T(\theta)\approx\left(\begin{array}{l}
        \cos{\theta}\\
        (\beta\pm i \alpha)\sin{\theta}\\
    \end{array}\right),
\end{align}
for $\alpha,\beta \ll 1$, i.e.\ for $B_{x,y}\ll B_z$.
The principal behaviour of these functions across rf resonance is depicted in Fig.~\ref{fig:hcomps}(a). This spectral decomposition shows rf-resonant behaviour. The transverse field components are mapped onto the quadratures of the first harmonic according to $\beta\pm i\alpha\approx (B_x\mp i B_y)/B_z$ and with an oscillation amplitude proportional to $\sin{\theta}$. The latter is maximal for $\theta=\pi/2$, i.e.\ exactly on rf resonance. At the same time, the zeroth harmonic, i.e.\ the dc signal, exhibits dispersive behaviour that maps the total static field strength, which is proportional to $B_z$ in the first order approximation near resonance. This configuration represents a vector magnetometer, but the dc component of the signal is quite vulnerable to electronic and technical noise, which will limit this strategy in practice to sensitive measurements of only the two transverse field components. 

When measuring the Voigt effect, the spectral decomposition of the signal leads up to the second harmonic and is given by
\begin{equation}
\Braket{\hat{S'}_z(t)}=\frac{1}{2}G_F^{(2)}S_y n_F\hbar^2\xi_F(F_z)\sum_{n=0}^2 h_n(\theta)e^{in\omega t}+c.c.,
\label{eqn:spectraloutput}
\end{equation}
using $\Bra{F,F_z}\hat{F}_y^2-\hat{F}_z^2\Ket{F,F_z}=\hbar^2(F(F+1)-3F_z^2)/2=\hbar^2\xi_F(F_z)$. 

According to Eqs.~(\ref{eq:h0Fxy})-(\ref{eq:h2Fxy}), the spectral components in the small angle approximation are in this case 
\begin{align}
    \label{eqn:approxVoigtharm}
    &\left(
    h_0,h_1,h_2\right)^T(\theta)\approx \left(\begin{array}{l}
        0\\
        {(\beta\mp i\alpha)\sin{2\theta}}
        \\
        {-\sin^2{\theta}} \\
    \end{array}\right).
\end{align}

The principal behaviour of these functions is shown in Fig.~\ref{fig:hcomps}(b). Again, the transverse field components are mapped onto the quadratures of the first harmonic, but now with a dispersive shape given by an oscillation amplitude proportional to $\sin{2\theta}$. Maximal amplitude is reached at $\theta=\pi/4$ and $\theta=3\pi/4$, i.e.\ when the static field is $B_\mathrm{dc}=B_\mathrm{sense}^\pm=B_{\mathrm{res}}\pm B_\rho$. In contrast to the Faraday decomposition, the zeroth harmonic vanishes while the second harmonic depends on the static field amplitude. Conveniently, the maximum sensitivity and approximately linear response to $B_z$ is also met at $B_\mathrm{dc}=B_\mathrm{sense}^\pm$. Hence, the Voigt rotation enables low-noise detection of all three magnetic components by evaluating the first and second signal harmonics.
For the Voigt effect measurements presented in the following, we work on the high field side of the rf resonance, by applying a field in the $z$-direction of strength $B_\mathrm{offs}=B_\mathrm{sense}^+(\alpha=\beta=0)=B_\mathrm{res}+B_\mathrm{rf}/2$. At this setting, the explicit second order expansion of the three relevant signal quadratures is given by
\begin{align}
\Re(h_1)=h_x=&+\left(\frac{1}{B_\mathrm{offs}} - \frac{\delta B_z}{B_\mathrm{offs}^2} \right)B_x, \label{Re1f}\\
\Im(h_1)=h_y=&-\left(\frac{1}{B_\mathrm{offs}} - \frac{\delta B_z}{B_\mathrm{offs}^2} \right)B_y, \label{Im1f}\\
\Re(h_2)=h_z=&-\frac{1}{2}+\frac{\delta B_z}{B_{\mathrm{rf}}}- \left(\frac{\delta B_z}{B_{\mathrm{rf}}}\right)^2\nonumber\\
&+\frac{2B_\mathrm{offs}+B_\mathrm{rf}}{4B_\mathrm{offs}^2B_\mathrm{rf}}B_x^2+\frac{B_\mathrm{offs}+B_\mathrm{rf}}{2B_\mathrm{offs}^2B_\mathrm{rf}}B_y^2\label{Re2f}.
\end{align}

\section{\label{sec:expColdatoms}Experimental realization: Laser cooled atoms}

\subsection{Laser cooled atoms setup}

Our experimental cold atom setup was described in~\cite{Jammi}, and here we will present only a brief description. We prepare an ensemble of approximately $2\times10^7$ laser cooled $^{87}$Rb atoms with a temperature of $(80\pm10)~\mathrm{\mu K}$ in the $\ket{F=2, m_F=0}$ state, using a sequence of optical pumping and state cleaning steps. Here, optical pump light propagating along a second optical axis was used to simplify the state preparation.
Atoms are then adiabatically dressed with a magnetic rf field in the $x$-direction with frequency $\omega = 2\pi \times 180~\mathrm{kHz}$, generated by an external resonant coil. The rf field amplitude is ramped up to $\approx 15~\mathrm{mG}$ over $4~\mathrm{ms}$ while the static magnetic field is ramped to a magnitude of $B_{\mathrm{offs}} \approx 260~\mathrm{mG}$ along the $z$-direction, which tunes the atomic Larmor frequency near resonance. 

\begin{figure}[b]
\centering
\includegraphics[width=0.45\textwidth]{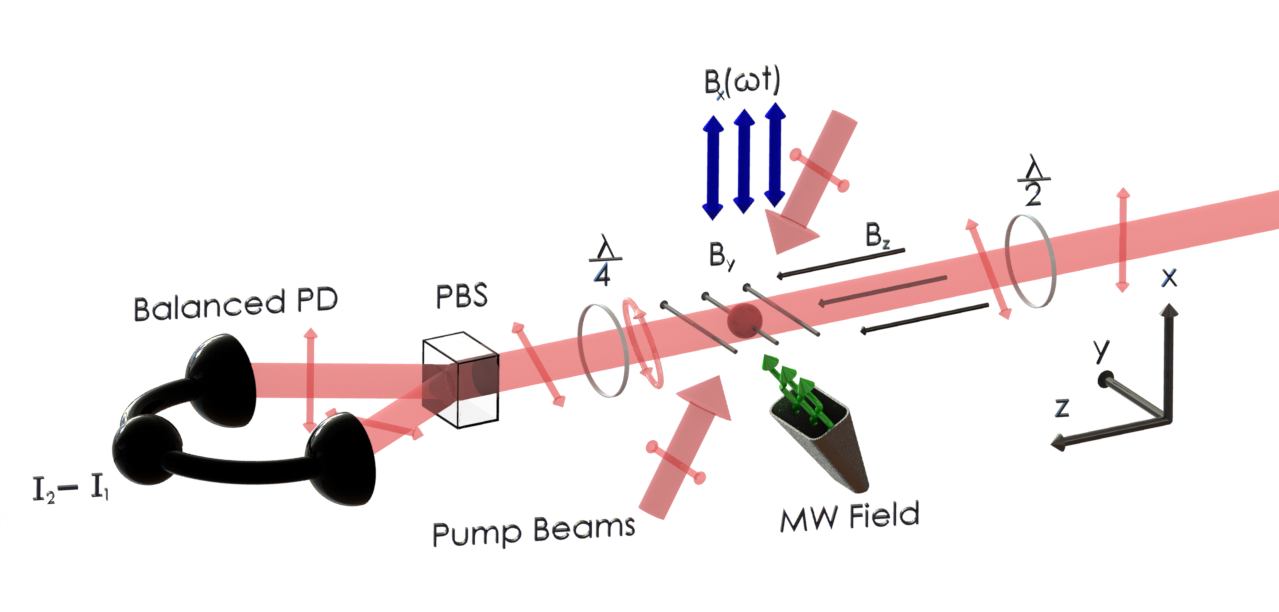}
\caption{Experimental setup. A laser cooled $^{87}$Rb sample is prepared in a pure $|F=2,F_y=0\rangle$ state. After rotation of the static field into the z-direction and adiabatic dressing with a magnetic rf field along x, linear birefringence of the sample is probed polarimetrically by a laser pulse propagating along z.}
\label{fig:sExpSetup}
\end{figure}

We measure the Voigt effect with a laser beam ($P\approx 100~\mu\mathrm{W}, \diameter\approx2.5~\mathrm{mm}$) detuned by $-400~\mathrm{MHz}$ from the $F=2\to F'=2$ transition of the $^{87}$Rb-$D_1$-line. A half-waveplate sets the polarization at $45\degree$ with respect to the $x,y$-axes. After interaction of a $1~\mathrm{ms}$ long probe pulse with the ensemble, a quarter-wave plate and a Wollaston prism allow us to measure the linear birefringence of the medium. The light is detected on a balanced photodetector pair (Thorlabs PDB210A) with a high-pass filtering rf amplifier (Minicircuits Model ZFL-1000+). The output voltage $u$ is proportional to the observed ellipticity, i.e.\ $u(t)=g_{\mathrm{el}} S'_z(t)$ with electronic gain $g_{\mathrm{el}}$, on the order of $10^{-12}~$V/Hz. The output signal is acquired by a field programmable gate array (FPGA) and is demodulated digitally with reference to the phase of the rf field.

\subsection{\label{sec:fieldmapping}Field mapping with laser cooled atoms}

The atomic ensemble should operate as a vector magnetometer near $\mathbf{B}=B_\mathrm{sense}^+\mathbf{e}_z$, where the two frequency modes of the atomic response map the three components of the magnetic field. Detuned to one HWHM above the rf resonance, the signal amplitude at $2\omega$ will be sensitive to the longitudinal field, while the two quadratures of the signal amplitude at $\omega$ map the transverse fields.

\begin{figure}[t!]
\begin{overpic}[width=0.45\textwidth]{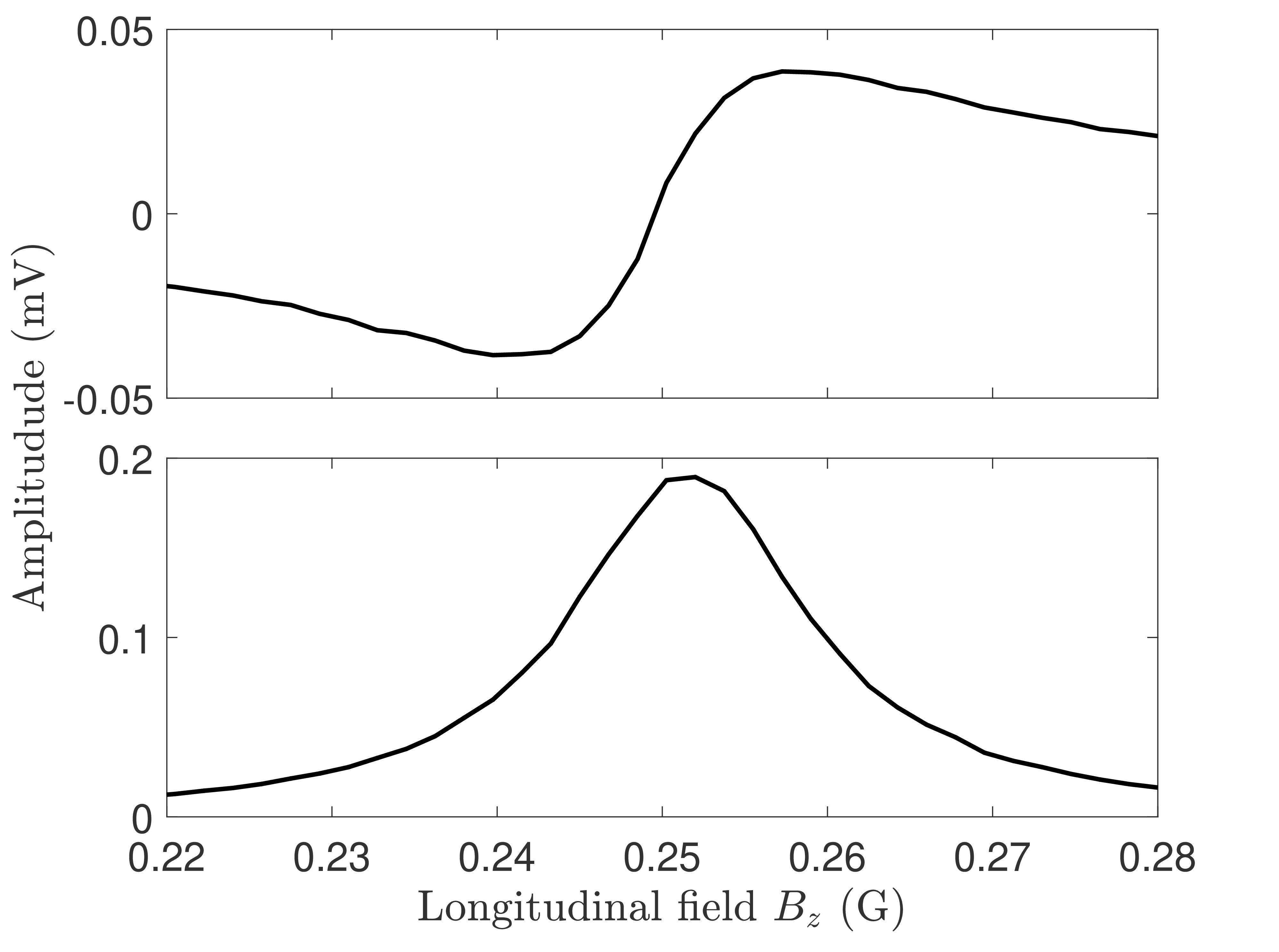}
\put (15,68) {a)} \put (15,35) {b)}
\end{overpic}
\caption{Voigt effect measurement across RF resonance. Typical, experimental amplitudes of the signal harmonics. Here, in-phase components of ac voltage amplitudes oscillating at $\omega$ (a), and at $2\omega$ (b) vary when the field component $B_z$ is scanned across the rf resonance in the presence of a constant, non-zero transverse field.}
\label{fig:h1h2ColdAtoms}
\end{figure}

\begin{figure}[t!]
\begin{overpic}[width=\textwidth/2]{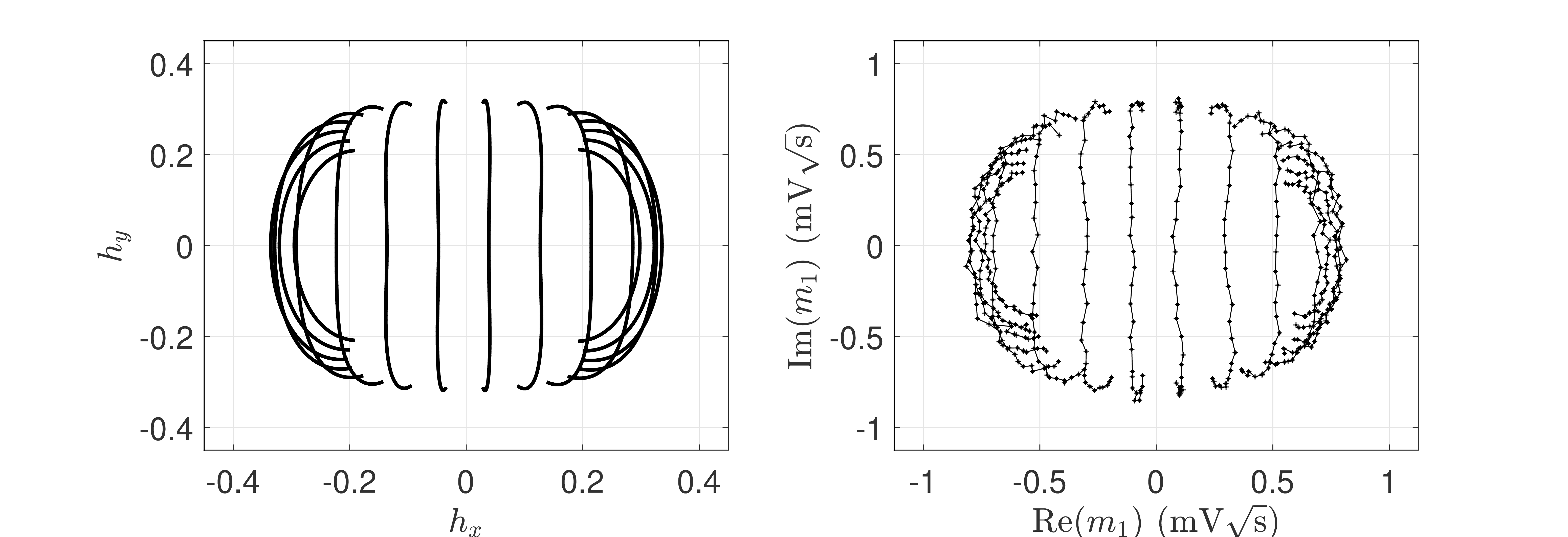}
\put(0,33){a)} \put(50,33){c)}
\end{overpic}\vspace{0.5cm}
\begin{overpic}[width=\textwidth/2]{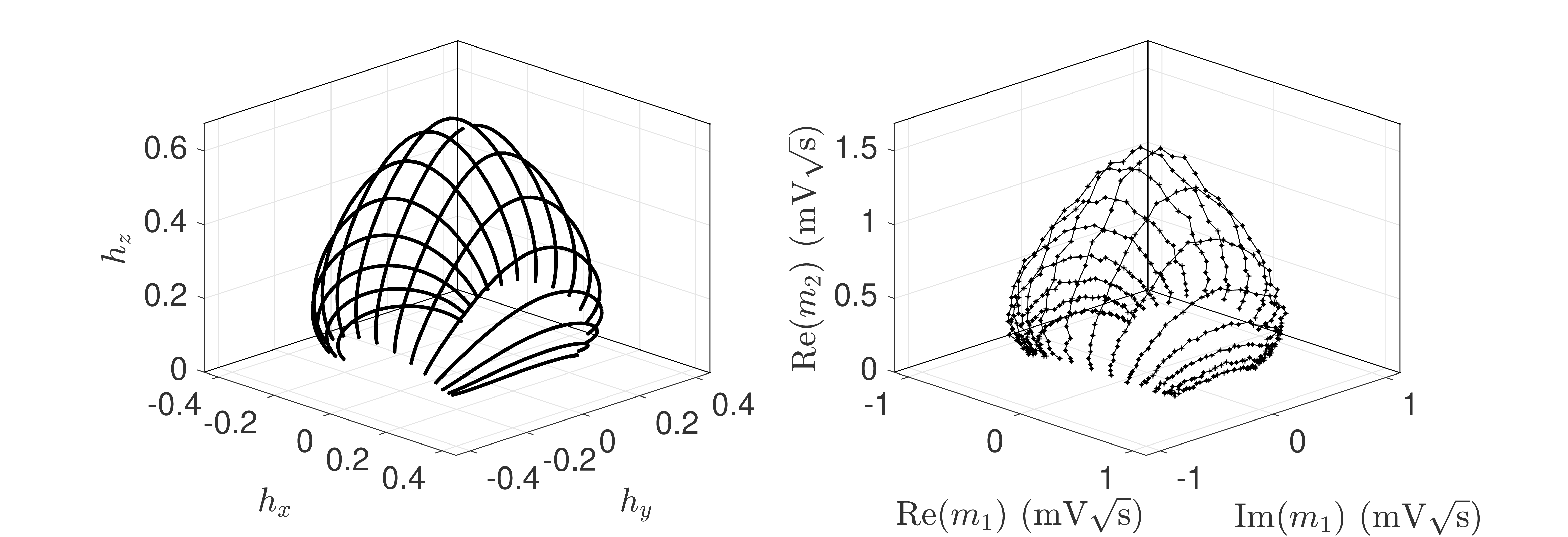}
\put(0,33){b)} \put(50,33){d)}
\end{overpic}
\caption{Atomic response, mapping the external magnetic field vector onto harmonic signal components. (a) Theoretical response of the signal quadratures at frequency $\omega$, as a function of the transverse field components $B_x$ and $B_y$ for constant $B_z$. (b) Full 3-dimensional response (see Appendix), including the real part of the signal amplitude at frequency $2\omega$, which allows for the measurement of the longitudinal field $B_z$. (c) Experimental realization, showing real and imaginary parts of the complex signal amplitude $m_1$ (detailed description in Sec.~\ref{subsec:detection}) at $\omega$ as a function of scanned transverse fields. The separation between two vertical lines is $\Delta B_x\approx 9~\mathrm{mG}$. 
(d) Full experimental response of both amplitudes $m_1$ and $m_2$ as a function of the scanned field.}
\label{fig:GridColdAtoms}
\end{figure}

Figure~\ref{fig:h1h2ColdAtoms} shows experimental signals demodulated at frequencies $\omega$ and $2\omega$ as a function of the static field $B_z\mathbf{e}_z$ (see more details in Section~\ref{subsec:detection}). The detected oscillation amplitude at frequency $2\omega$ shows resonant behaviour. The dispersive responses at frequency $\omega$ are observed due to the presence of a transverse field with non-zero $x$ and $y$ components. On the high field side, these signals show maximal amplitude near $B_\mathrm{z}=B_\mathrm{sense}^+\approx 0.258~\mathrm{G}$, where the $2\omega$ amplitude shows an approximately linear response with respect to $B_z$. To show vector magnetometer operation, we scan the transverse fields in a grid like pattern at constant $B_z$ near the sensitive point. The results are shown in Fig.~\ref{fig:GridColdAtoms}, together with the matching theoretical response.

The results confirm the principle of operation for the magnetometer based on the Voigt effect in cold atoms, with well controlled preparation of pure quantum states and temporal separation of state-preparation and exposure to an external field. For practical purposes, this setup is of limited use, given the complexity of the apparatus and limitations on achievable sample proximity and bandwidth or cycle rate. Therefore, we explored Voigt effect magnetometry in a vapour cell with room temperature atoms towards practical devices with higher bandwidth and sensitivity.

\section{\label{sec:expvapour}Experimental realisation: Room temperature vapour}

\subsection{Room temperature vapour setup}

\begin{figure}[b]
\begin{overpic}[scale=0.3]{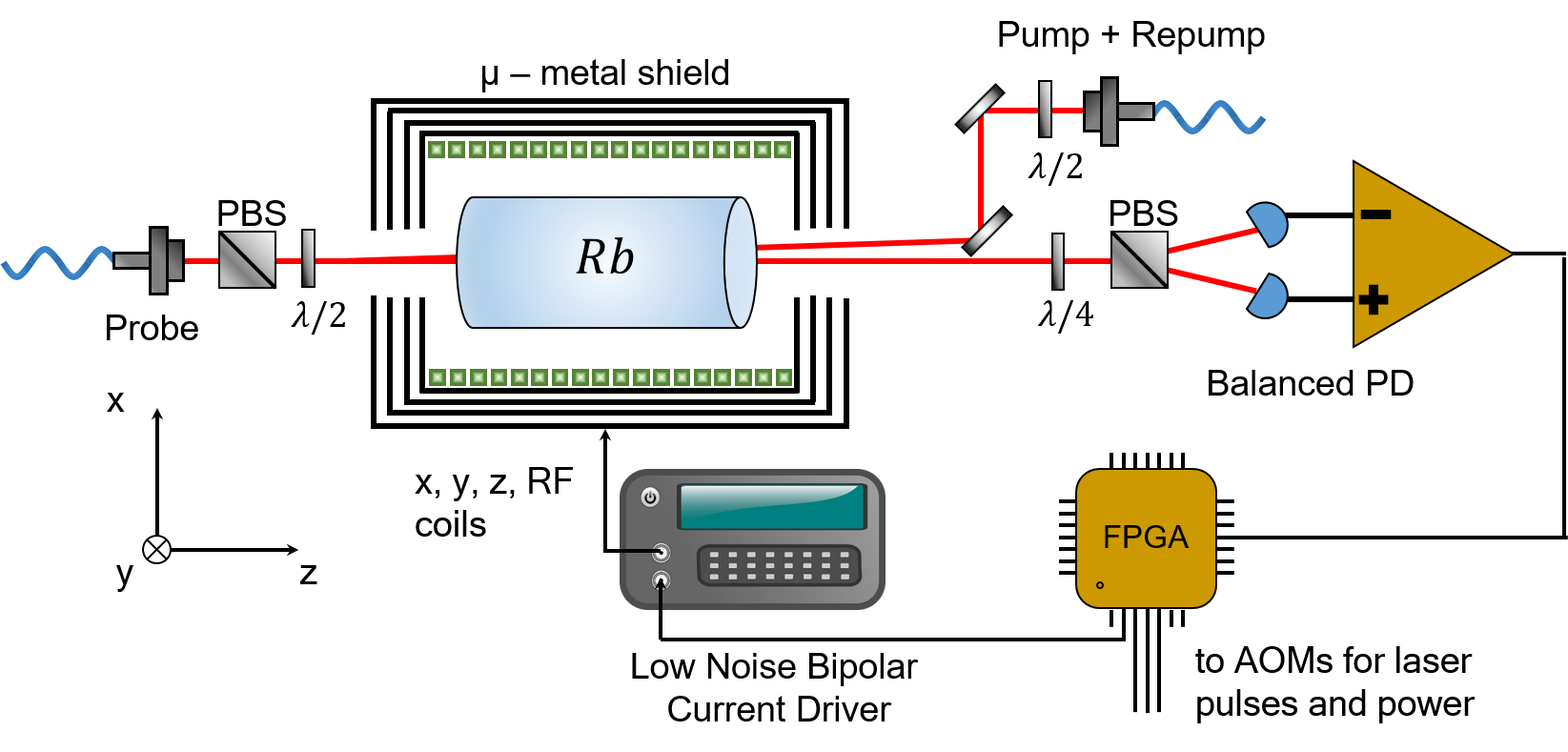}
\put(0,43){a)}
\end{overpic}\quad
\begin{overpic}[scale=0.22]{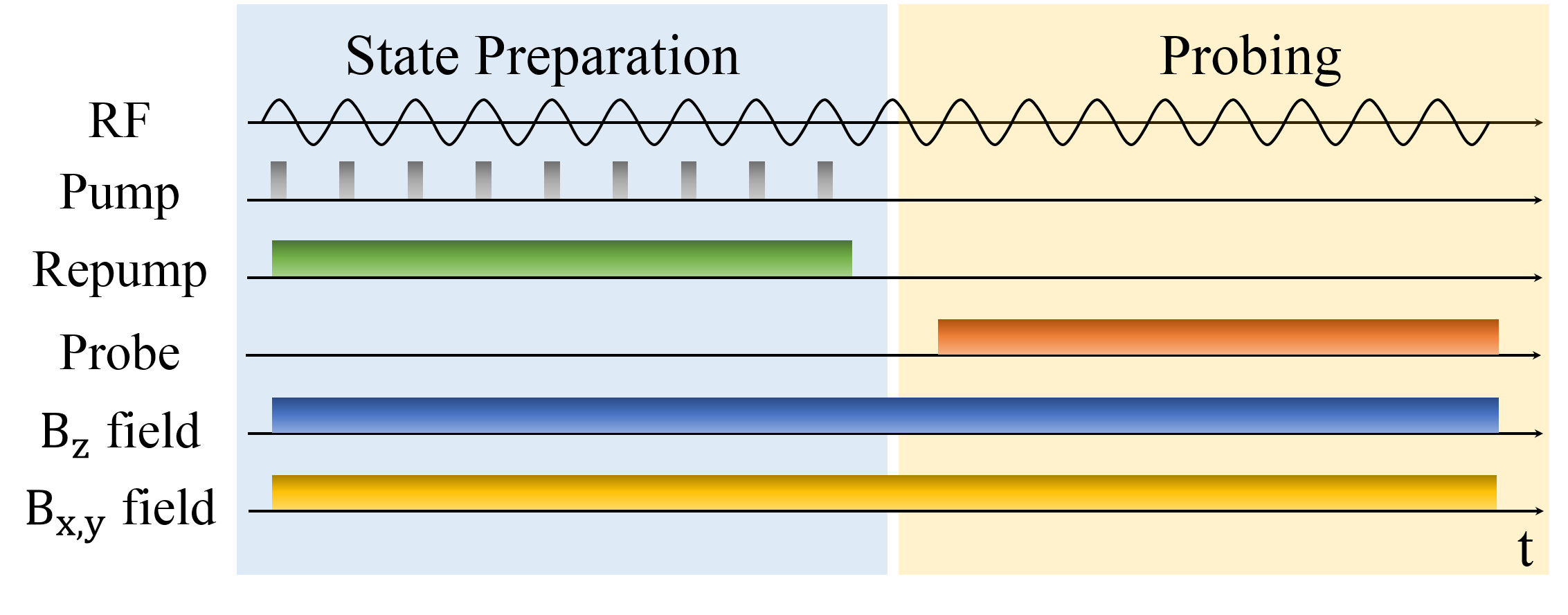}
\put(0,40){b)}
\end{overpic}
\caption{Experimental realization with a shielded vapour cell. (a) Layout of the experimental setup. A pair of pump beams is used to prepare the atomic state before a polarimetric measurement is performed on a counterpropagating probe beam. The pump light is incident under a small angle with respect to the probe and the static offset field. (b) Single shot experimental pulse sequence with a typical 8 ms duration. During each cycle, the static fields are held at constant, scanned values after switching within $\leq 50~\mu\mathrm{s}$ at the start of the state preparation process.}\label{fig:setup}
\end{figure}

Our setup is based on a paraffin coated $^{87}$Rb enriched vapour cell of diameter $d=26$ mm and length $l=106$ mm at room temperature, with a density of approximately $10^{10}$ atoms per cubic cm.
The cell is placed inside a commercial 4-layer $\mu$-metal shield (Twinleaf MS-2), mounted on a non-magnetic vibration isolation table with non-magnetic optomechanics, see Fig.~\ref{fig:setup}(a). The static magnetic fields and the radio frequency field inside the chamber are generated by a combination of a solenoid for the longitudinal field and cosine-theta coils for the transversal fields. The coils are driven by a lead-acid battery powered, ultra low-noise current sources, based on the modified Hall-Librecht design \cite{hall,dalin}. The laser system to address the atomic transitions for state preparation and probing consists of a combination of commercial and in-house external cavity diode lasers. The laser system is housed on a separate vibration isolation table, and the light is coupled via single mode, polarization maintaining fibers. 

The atomic vapour is optically pumped with a linearly polarized pair of laser beams, counterpropagating to a linearly polarized probe beam. A small angle is used for optical access.  
The Voigt rotation is measured by separating the two circular polarization components with a quarter wave-plate and polarizer cube and detecting the light with a balanced photodetector pair (Thorlabs PDB210A), with electronic gain $g_{\mathrm{el}}\approx10^{-13}~$V/Hz. The magnetometer is operated in pump-probe mode, where each cycle contains an initial period of synchronous optical pumping before probing the atomic state. The experimental sequence generation and data acquisition are performed using a National Instruments FPGA (PCIe-7852).

\subsection{\label{subsec:state_preparation}State preparation}

In each cycle, we perform the state preparation by optical pumping, which reaches a steady state over the first 5~ms, before probing the state for another 3~ms, see Fig.~\ref{fig:setup}(b). 
But different from the cold atom test case, the magnetic fields are not adiabatically ramped to different values between pump and probe stages. Here, we prepare dressed states directly by synchronous pumping \cite{sync_pump}, i.e.\ using a pulse train of pump light, in phase with a uniform, 5~kHz rf field of $\lessapprox 0.1$ mG, near resonant with the static field $B_z$.
We use linear polarization along the $x$-axis, parallel to the rf field. During a short 9\% duty cycle, this direction is nearly aligned with our quantization axis in the rotating frame. This enables the preparation of either the dressed state $\ket{F=2,m_F=0}$ for pumping near the $F=2\rightarrow F'=2$ transition on the $D_1$ line, or an incoherent mixture of $\ket{F=2,m_F=\pm2}$ for pumping near $F=2\rightarrow F'=1$. Our signal amplitude depends quadratically on the magnetic quantum number $m$, and all three of these states give rise to the same maximally possible signal with $\xi_F(2)=\xi_F(-2)=-\xi_F(0)$.
The choice of states and required pump polarization allows for the pump beam to propagate parallel to the probe beam, along the $z$-axis. 

\begin{figure}[t!]\centering\hspace*{-1em}
\begin{overpic}[width=\textwidth/2]{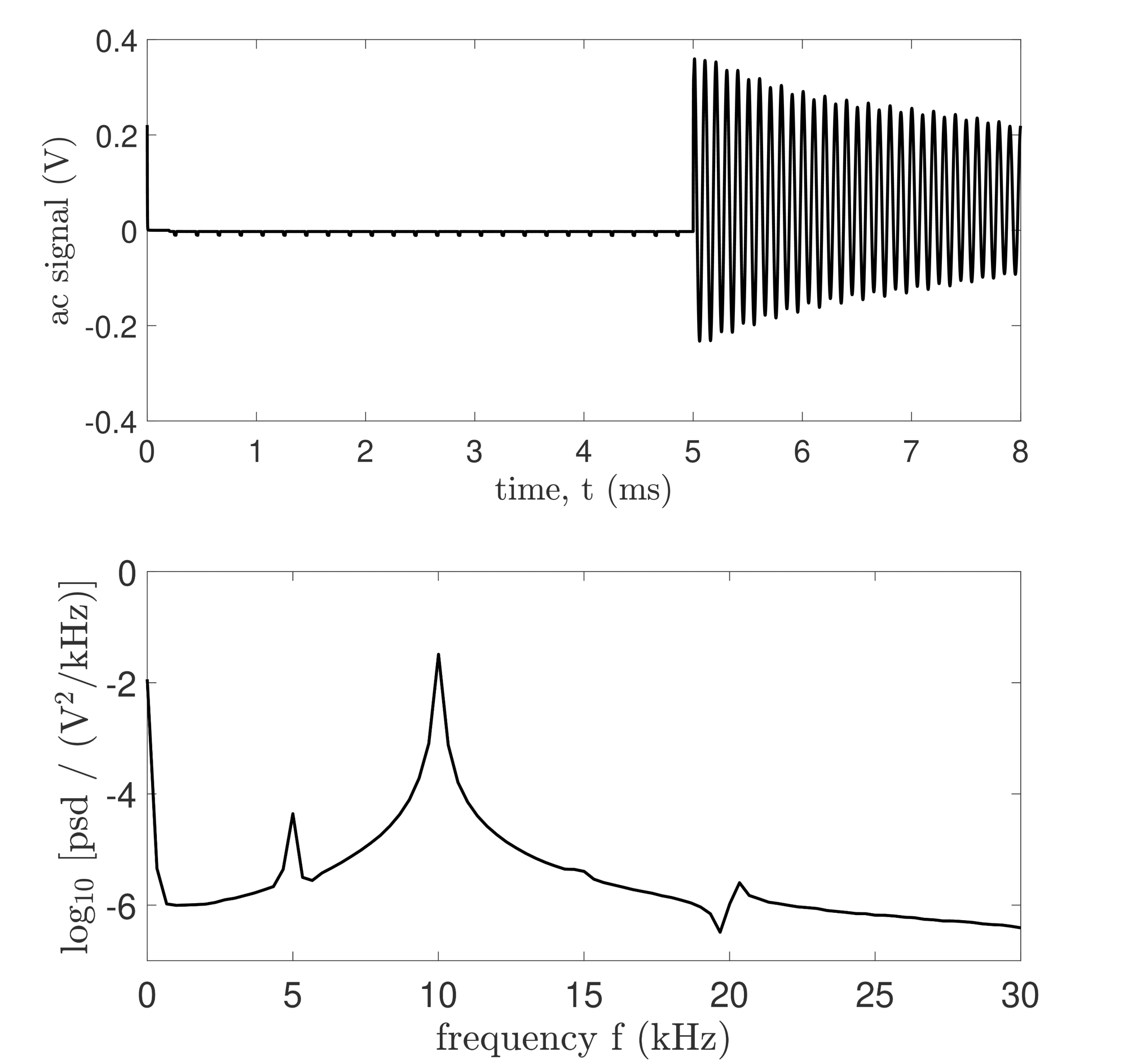}
\put (3,90) {a)} \put (3,45) {b)} 
\end{overpic}
\caption{Typical experimental signals. a) Raw balanced signal of the Voigt rotation. The state preparation occurs within the first 5ms of the cycle followed by a 3ms probing pulse. b) Single-sided, power spectral density (psd) of the amplified signal during the probe pulse. Atomic signals arise at $\omega$ and $2\omega$. Weak harmonics at $3\omega$ and $4\omega$ can also be observed, which may arise due to non-linear magneto-optical effects~\cite{Budker2002} and non-linearities in the electronic detection path. For comparison, photon shot noise is at a level of $g_\mathrm{el}^2S_y\approx 0.6\times10^{-10}~\mathrm{V}^2/\mathrm{kHz}$.
}\label{fig:raw_data}
\end{figure}
\begin{figure}[h!]\vspace*{-0.3cm}\hspace*{-0.1cm}\centering
\begin{overpic}[width=\textwidth/2]{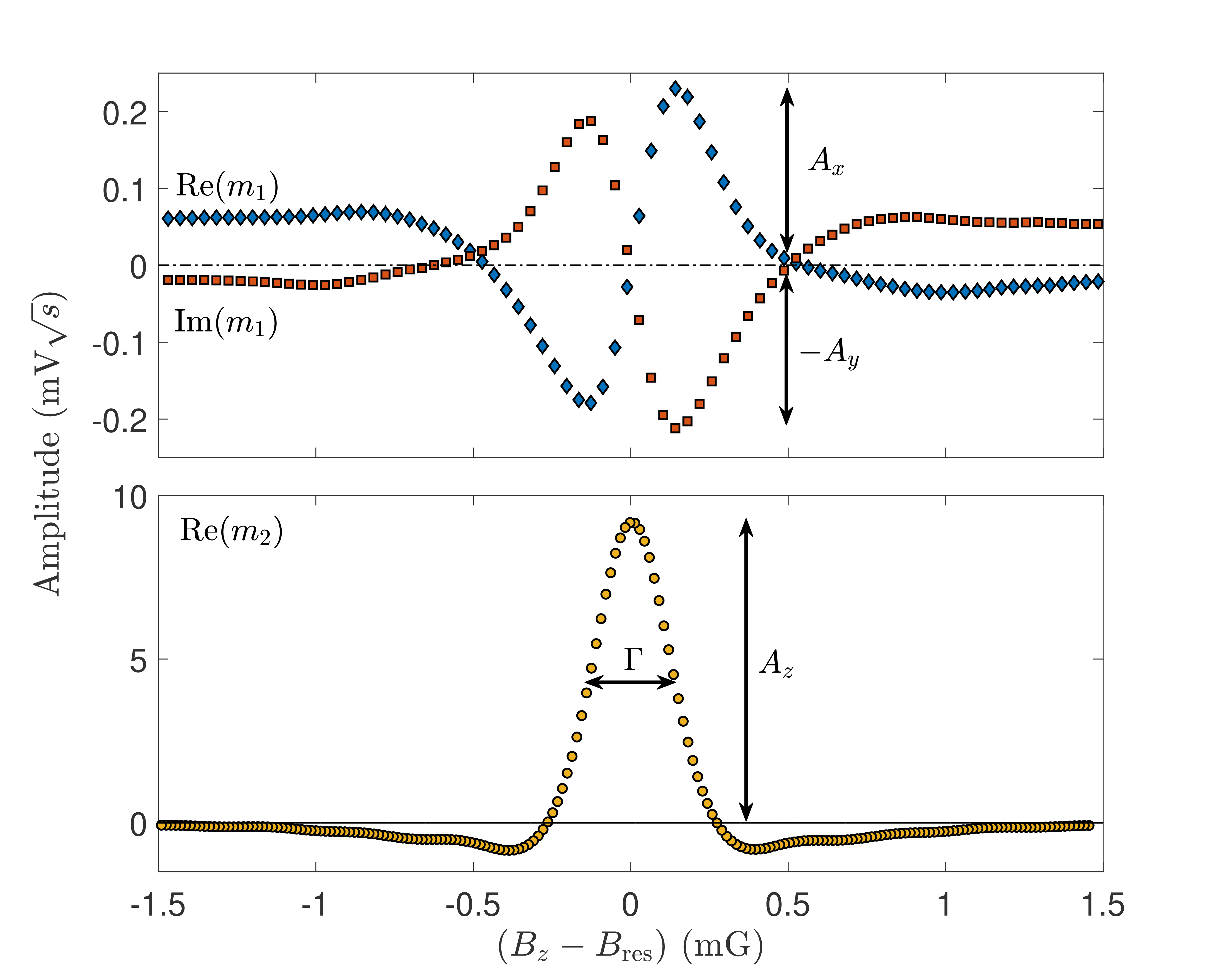}
\put (-2,73) {a)} 
\put (-2,39) {b)} 
\end{overpic}
\hspace*{-0.1cm}\centering\begin{overpic}[width=\textwidth/2]{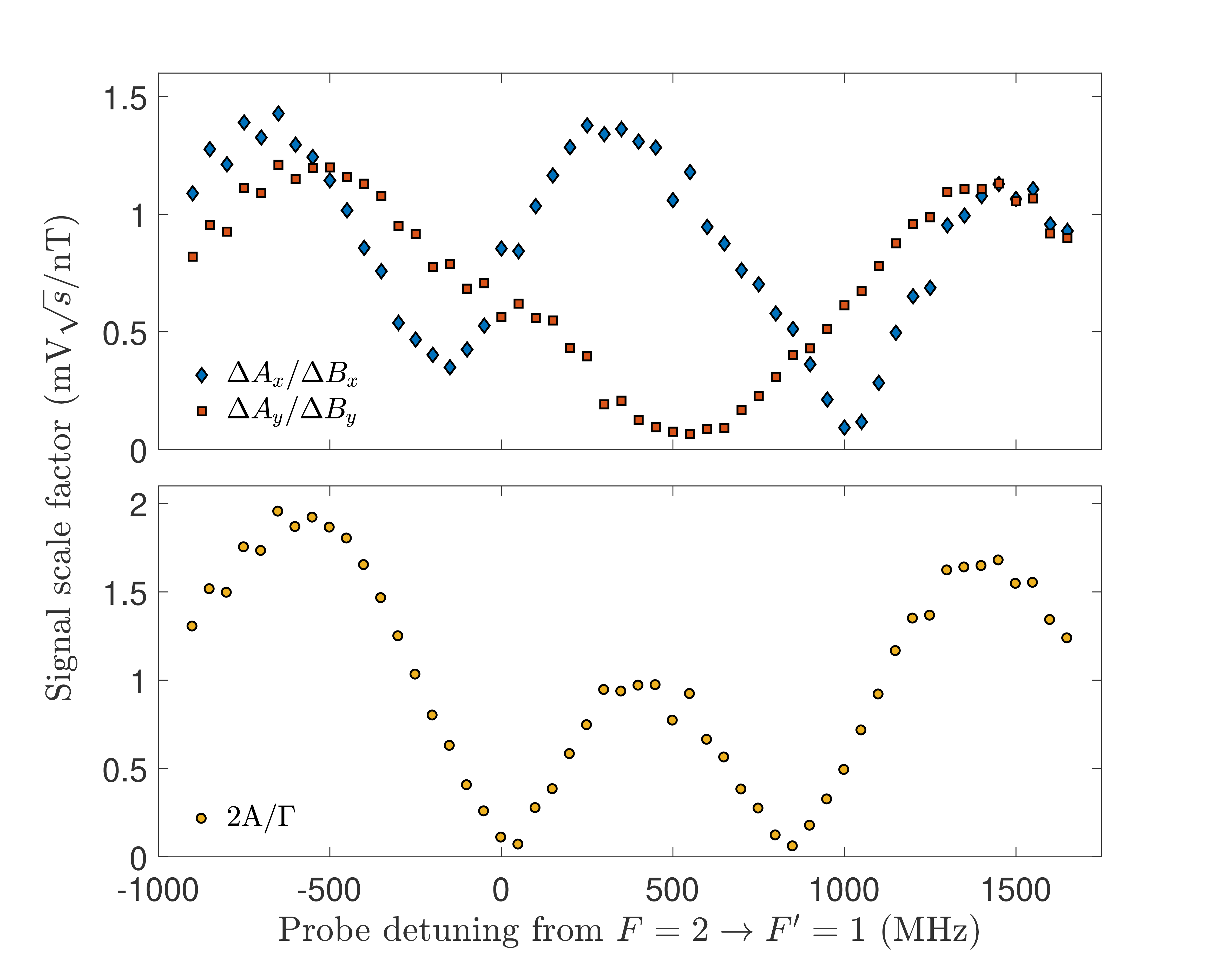}
\put (-2,74) {c)}
\put (-2,41) {d)}
\end{overpic}
\caption{Experimental magnetometer response. Panels a) to b) show the three relevant quadratures of the mode amplitudes $m_1$ and $m_2$, i.e.\ responses at frequencies $\omega$ and $2\omega$, for a scan of the longitudinal field $B_z$ across the rf resonance. Here, non-zero transverse fields $B_x$ and $B_y$ are kept constant. The mode amplitudes, extracted according to Eq.~(\ref{eq:mode_amplitudes}), follow the predicted behaviour, see Eq.~(\ref{eq:Voigt}). Panels c) and d) show experimental estimates for the three signal scale factors as a function of probe detuning. For the longitudinal field, this is the slope of the $2\omega$ resonance profile, estimated as $2A/\Gamma$. Near the chosen probe detuning of -550MHz, all scale factors are close to maximal, and the first order responses to orthogonal external fields are orthogonal.}
\label{fig:opm_1f2fmodes}
\end{figure} 

To maximise atomic population in the $F=2$-hyperfine manifold, a co-propagating CW repump beam addressing $\ket{F=1}\rightarrow \ket{F'=2}$ of the D2 line of the same polarization is spatially overlapped with the pump, which repopulates atoms from the $\ket{F=1}$ to the $\ket{F=2}$ ground state. The pump and repump beams share the same Gaussian intensity profile with 7.3~mm diameter ($1/e^2$) and 2.2 mW/$\mathrm{cm^2}$ and 1.6 mW/$\mathrm{cm^2}$ peak intensity, respectively. The pumping efficiency is limited by atom exchange, spin exchange collisions, other decoherence processes, non-parallel effective field and pump polarisation at the sensitive field offset $B^+_{\mathrm{sense}}$ where $\theta=\pi/4$, and the synchronized pump duty cycle, which is a compromise between effective power and achieving momentary alignment between effective field and polarisation.
The resulting spin state can be characterized spectroscopically~\cite{Julsgaard}, and here we used a stroboscopic version of microwave spectrocopy to probe the dressed atomic states~\cite{Sinuco}. The experimental lower estimate confirmed that more than 75\% of the atomic population is pumped into a mixture of the dressed states $\ket{F=2,m=\pm2}$. Details of this method will be published elsewhere. The imperfect pumping efficiency reduces the overall signal strength due to a reduction of the atomic alignment but may also influence collisional dynamics at sufficiently high atomic densities.

\subsection{\label{subsec:detection}Signal detection}
Immediately after the state preparation process we couple a counter-propagating probe pulse along $z$ to measure the resultant Voigt rotation. The probe is detuned by -550 MHz from the $\ket{F=2}\rightarrow \ket{F'=1}$ transition of the D1 line. It has a Gaussian profile of 3.4~mm diameter ($1/e^2$) and 2.6~mW/$\mathrm{cm^2}$ peak intensity and linear polarization set to 45\textdegree with respect to the rf field. 

\begin{figure*}[t]
\begin{overpic}[width=\textwidth]{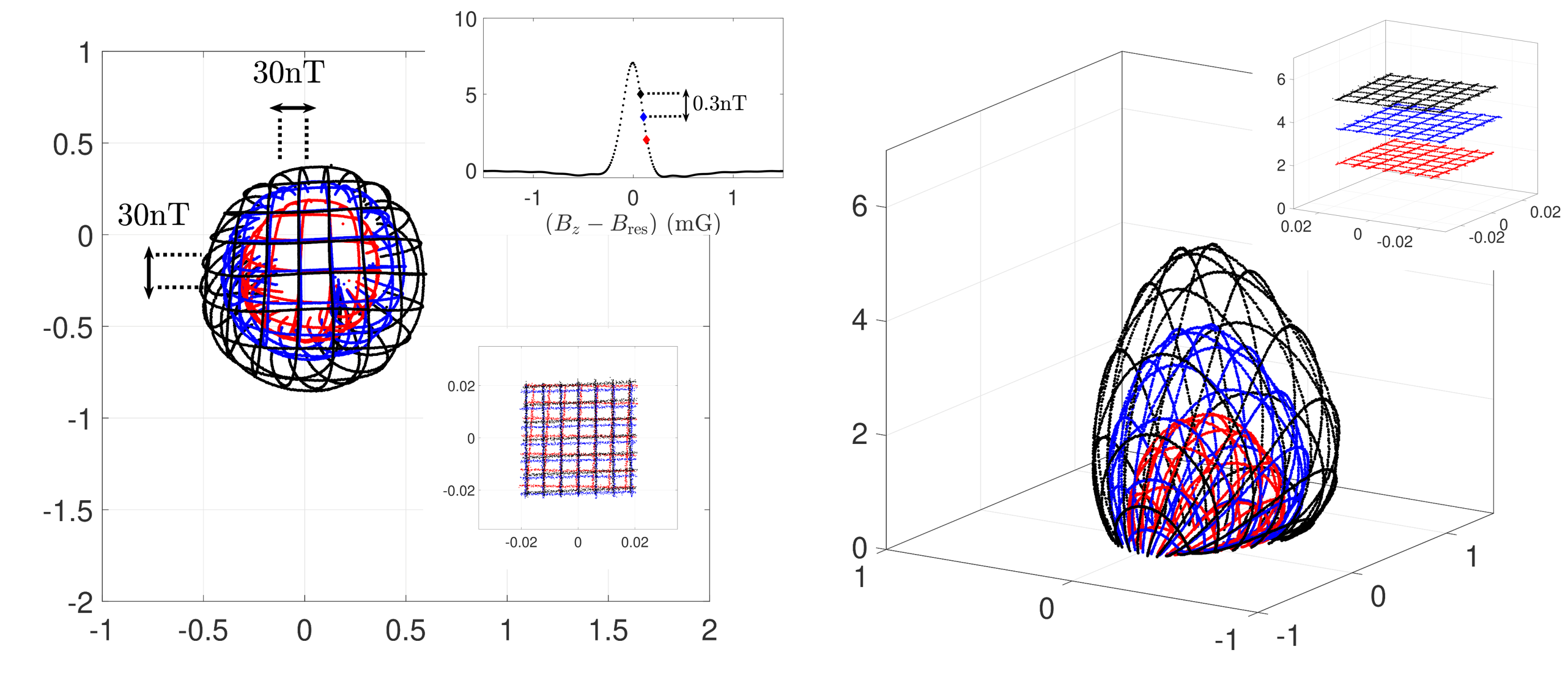}
\put (2,40) {a)} 
\put (54,40) {b)} 
\put (0,18) {\rotatebox{90}{Im($m_1$) (mV$\sqrt{\mathrm{s}}$)}} 
\put (21,1) {Re($m_1$) (mV$\sqrt{\mathrm{s}}$)} 
\put (27,32.5) {\rotatebox{90}{\scriptsize{Re($m_2$) (mV$\sqrt{\mathrm{s}}$)}}}
\put (26.5,10.5) {\rotatebox{90}{\scriptsize{Im($m_1$) (mV$\sqrt{\mathrm{s}}$)}}}
\put (31.5,7.5) {{\scriptsize{Re($m_1$) (mV$\sqrt{\mathrm{s}}$)}}} 
\put (51,16) {\rotatebox{90}{Re($m_2$) (mV$\sqrt{\mathrm{s}}$)}}
\put (60,3.5)
{\rotatebox{-8.5}{Im($m_1$) (mV$\sqrt{\mathrm{s}}$)}}
\put (83,1)
{\rotatebox{25}{Re($m_1$) (mV$\sqrt{\mathrm{s}}$)}}
\put (79,29) {\rotatebox{90}{\scriptsize{Re($m_2$) (mV$\sqrt{\mathrm{s}}$)}}}
\put (83,27.5) {\rotatebox{-8.5}{\scriptsize{Im($m_1$) }}}
\put (88,24.5) {\scriptsize{(mV$\sqrt{\mathrm{s}}$)}}
\put (94,26.2) {\rotatebox{26}{\scriptsize{Re($m_1$) }}}

\end{overpic}
\caption{Mapping of the OPM response to external fields at 5 kHz radio frequency dressing. (a) Quadratures of the first harmonic signal as a function of raster scanned transverse fields $B_{x,y}$ ranging over $\approx\pm180$~nT for constant $B_z$. Each colour represents a different $B_z$ field ranging over $\approx\pm0.3$~nT. The top inset shows the location of the $B_z$ field with respect to the resonant signal at $2\omega$. (b) Inclusion of the second harmonic signal produces oviforms in the three-dimensional representation and demonstrates the full vector mapping. The behaviour in the linear regime for small field perturbations is shown in the insets, with visible photon shot noise $g_\mathrm{el}\sqrt{S_y}\approx 2.5\times10^{-4}~\mathrm{mV}\sqrt{\mathrm{s}}$. We attribute deviations from the ideal profiles to geometric misalignment between the pump/probe beams and static and rf fields. The asymmetric distortion increases for lower bias fields, consistent with imperfect orthogonality between static field coils and their alignment with the direction of the probe beam.  
}
\label{fig:3d_plot}
\end{figure*}
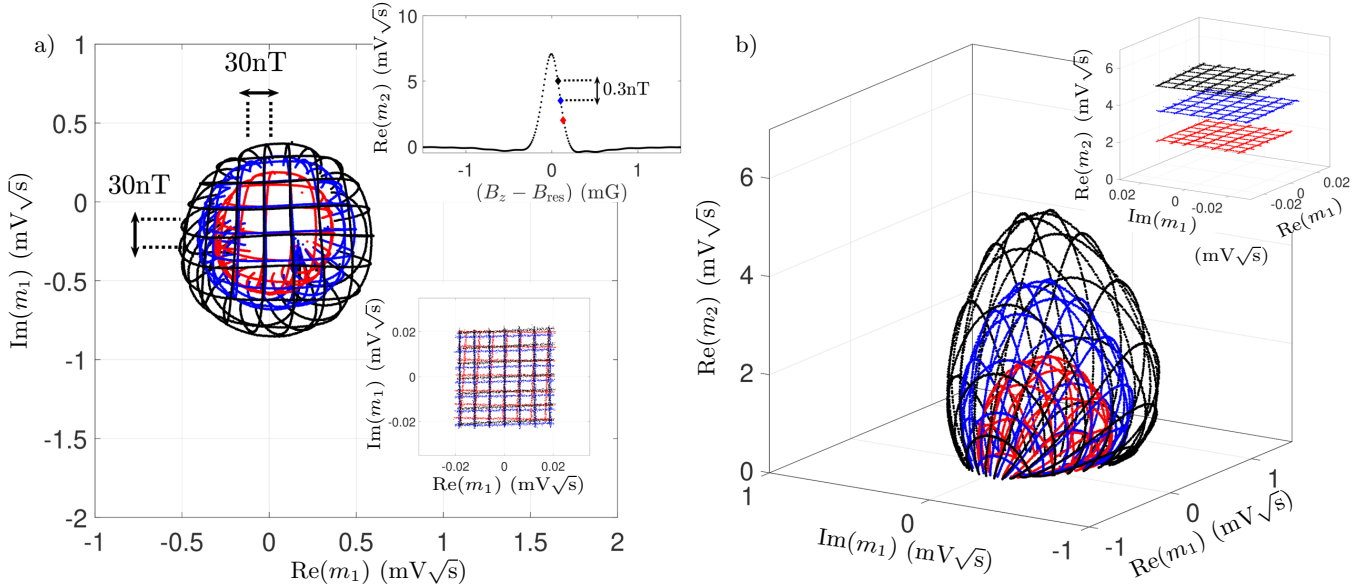

The interaction with the dressed atomic medium results in modulated, elliptical polarization of the probe. Figure~\ref{fig:raw_data} shows the typical temporal trace of the balanced detector signal during one cycle together with its spectrum. The main contributions to the rf signal are found at frequency 2$\omega=10$ kHz and a weaker signal at the dressing frequency $\omega=5$ kHz due to the presence of transverse fields. As can be seen in Fig.~\ref{fig:raw_data}(a), the atomic signal decays due to finite state lifetime. Generally, this is limited by the atom-wall and atom-atom collision rates. In our case, the exchange of the atoms between the main cell body and the stem with the Rb reservoir is the major contributing factor to the relaxation rate. In principle, high quality anti-relaxation coatings together with a lockable stem system can be used to achieve coherence lifetimes in excess of 60~s \cite{balabas}. 

Absorption of the probe beam introduces additional decay and additional optical pumping, which broadens the $2\omega$ resonance profile and alters the response at $\omega$ to transverse fields. We choose the combination of probe power and detuning by optimizing the slope of the $2\omega$-signal with respect to the external field strength. We evaluate this scale factor as the ratio of height $A_z$ and width $\Gamma/2$ of the resonance peak, both indicated in Fig.~\ref{fig:opm_1f2fmodes}(b). The dependence on probe detuning is shown in Fig.~\ref{fig:opm_1f2fmodes}(d). The maximal response to longitudinal fields is found away from the Doppler broadened absorption lines, where also the signal responses to small changes of the two orthogonal transverse field components are close to maximal, shown in Fig.~\ref{fig:opm_1f2fmodes}(s), and show the predicted $\pi/2$ relative phase shift. Closer to the resonances, we observe different and non-orthogonal responses.

The detected raw signals are digitally demodulated to extract the three-dimensional field vector information. We calculate complex temporal mode amplitudes $m_k$ for the first and second harmonic ($k=1,2$) by taking the scalar product of the signal $u(t)=g_\mathrm{el}S'_z(t)$ with exponentially decaying, normalised mode functions leading to the definition
\begin{equation}
m_k=\int_{t_1}^{t_2}e^{(-ik\omega-\gamma)t}u(t)dt/\sqrt{\int_{t_1}^{t_2}e^{-2\gamma t}dt},\label{eq:mode_amplitudes}
\end{equation}
which covers the interval of the probe pulse between times $t_1$ and $t_2$ and matches the atomic response. To first order and for appropriately adjusted phase, the quadratures, i.e.\ the real and imaginary part of $m_1$ reflect the external field components $B_x$ and $B_y$ while the real $m_2$ is sensitive to $B_z$. 
The signals are phase-locked to the rf driving field, but they acquire additional electronic phase-shifts. Therefore, we first adjust the demodulation phase for the second harmonic signal by scanning the longitudinal $B_z$ field whilst the transverse fields are set to zero. We adjust the phase such that the real and imaginary quadratures of the mode amplitude $m_2$ produce a symmetric and a dispersive profile, respectively. Any additional phase entering the first harmonic is equivalent to a rotation of the field coordinate system about the longitudinal axis. Here, we scan the transverse fields over a small range, with the longitudinal field adjusted to the sensitive field point $B_z\approx B^+_{\mathrm{sense}}$ to identify two orthogonal quadratures with the $x,y$-transverse field coils by minimizing their crosstalk.
As it is shown in Fig.~\ref{fig:opm_1f2fmodes}(a) both quadratures of the signal amplitude $m_1$ follow a dispersive profile. Fig.~\ref{fig:opm_1f2fmodes}(b) shows the resonant response of the amplitude $\mathrm{Re}(m_2)$. This is consistent with the theoretical model described in section~\ref{sec:theory} and with the experimental results for cold atoms in section~\ref{sec:fieldmapping}. In contrast to the double resonance magnetometer described in Ref.~\cite{Weis1}, where the quadratures of the first harmonic present a resonant and a dispersive profile with respect to the applied static field, the quadratures in Fig.\ref{fig:opm_1f2fmodes}(a) are both dispersive while one quadrature of the second harmonic shows a resonant profile. All three signals become most sensitive to orthogonal external field components at the same offset field.

The conversion of measured signals into magnetic field values relies on a two-step calibration procedure. First, the static field coils are characterised with two independent methods, before the signal scale factor for each field direction is determined by applying a range of known fields to the magnetometer.
The static field coils together with their electronic drivers are calibrated using the known field dependence of the Larmor resonance for $^{87}$rubidium. For a set of fixed radio-frequencies $\omega$, the longitudinal field $B_z$ is scanned with no transverse fields present to find the maximal response of the $2\omega$ signal. Under this condition, the field is determined by known parameters according to $B_z=B_\mathrm{res}=\hbar\omega/\mu_B|g_F|$. The resonant field can be easily calculated and plotted against the applied voltage/current of the coils giving the field conversion. The presence of transverse fields changes the resonance condition to $B_\mathrm{res}=\hbar\omega/\mu_B|g_F|\sqrt{B_z^2+B_{x,y}^2}$. Thus, to obtain the calibration for the transverse fields, we change one of the transverse fields whilst keeping the other one at zero and sweep the $B_z$ field to obtain a new location of the $2\omega$ resonance. As before, the new resonance location is evaluated as a function of the control voltage/current of the coils. 
In addition to this procedure, we confirm the coil calibrations using a commercial fluxgate magnetometer (Stefan Mayer Instruments, FLC3-70).
Finally, the signal scale factors are measured by applying a linear field ramp of a well known range to each of the fields independently. For small fields, the corresponding demodulated signal responses show linear relationships, see Eqs~(\ref{Re1f})-(\ref{Re2f}), which we use to calibrate the signal-to-field conversion.
 
\subsection{\label{subsec:detection}3D vector mapping}

Following the same procedure as in the cold atom case in Section~\ref{sec:fieldmapping}, we measured the three components of the field by setting the static field $B_z=B_\mathrm{sense}$, which maximizes the mode amplitudes at $\omega$. By linearly scanning the external transverse fields and demodulating the $\omega$ and $2\omega$ quadratures, we are able to map the magnetometer response. The vector magnetometer operation can be visualized on a 3D plot shown in Fig.~\ref{fig:3d_plot}. The full mathematical description of the oviform plot can be found in the Appendix~\ref{sec:faraday}. But it must be noted that the theoretical results are based on the assumption of a pure atomic state. The model does not account for decoherence due to atomic collisions and various broadening effects (e.g. gradient fields, light power). Nevertheless, the experimental results are in reasonable agreement with the expected 3-dimensional response, see Eqs.~(\ref{Re1f})-(\ref{Re2f}). As can be seen in Fig.~\ref{fig:3d_plot} the 2D and 3D oviform profiles arising from magnetic field scans show some asymmetric distortion and an offset, which arise from geometrical misalignment of the probe/pump beams relative to the static fields and/or the rf dressing field as confirmed for larger misalignment. The insets show the magnetometer response in the linear regime for small external fields.
 
Despite the theoretical simplifications, our model effectively describes the vector magnetometer response to the external magnetic fields. 
We assume that effects of distortion can be reduced by accurate alignment between the probe/pump beams, the dressing field, and the small static offset fields. Operation at higher dressing frequency and thus larger offset field leads to more accurate alignment of the field across the atomic ensemble, however, this method reduces the sensitivity to the transverse fields as described in the following section. 
The sensing range for longitudinal field is determined by the width of the resonance, whilst the response to transverse fields and its non-linearity is determined by their ratio to the offset field.
In principle, the range of operation can be extended and maximal sensitivity maintained by placing the magnetometer into a closed loop system.
 
\subsection{\label{subsec:detection}Noise performance}

To perform the noise measurements we detune the static $B_z$ field to $B_z=B_\mathrm{sense}^+$, which optimizes the magnetometer sensitivity for all three components. We then adjust the transverse fields such that the first harmonic signal vanishes, i.e.\ the noise measurements are done near the apex of the middle (blue) ovoid in Fig.~\ref{fig:3d_plot}.(b).

\begin{figure}[ht!]\centering \hspace*{-0.3cm}
\begin{overpic}[width=\textwidth/2]{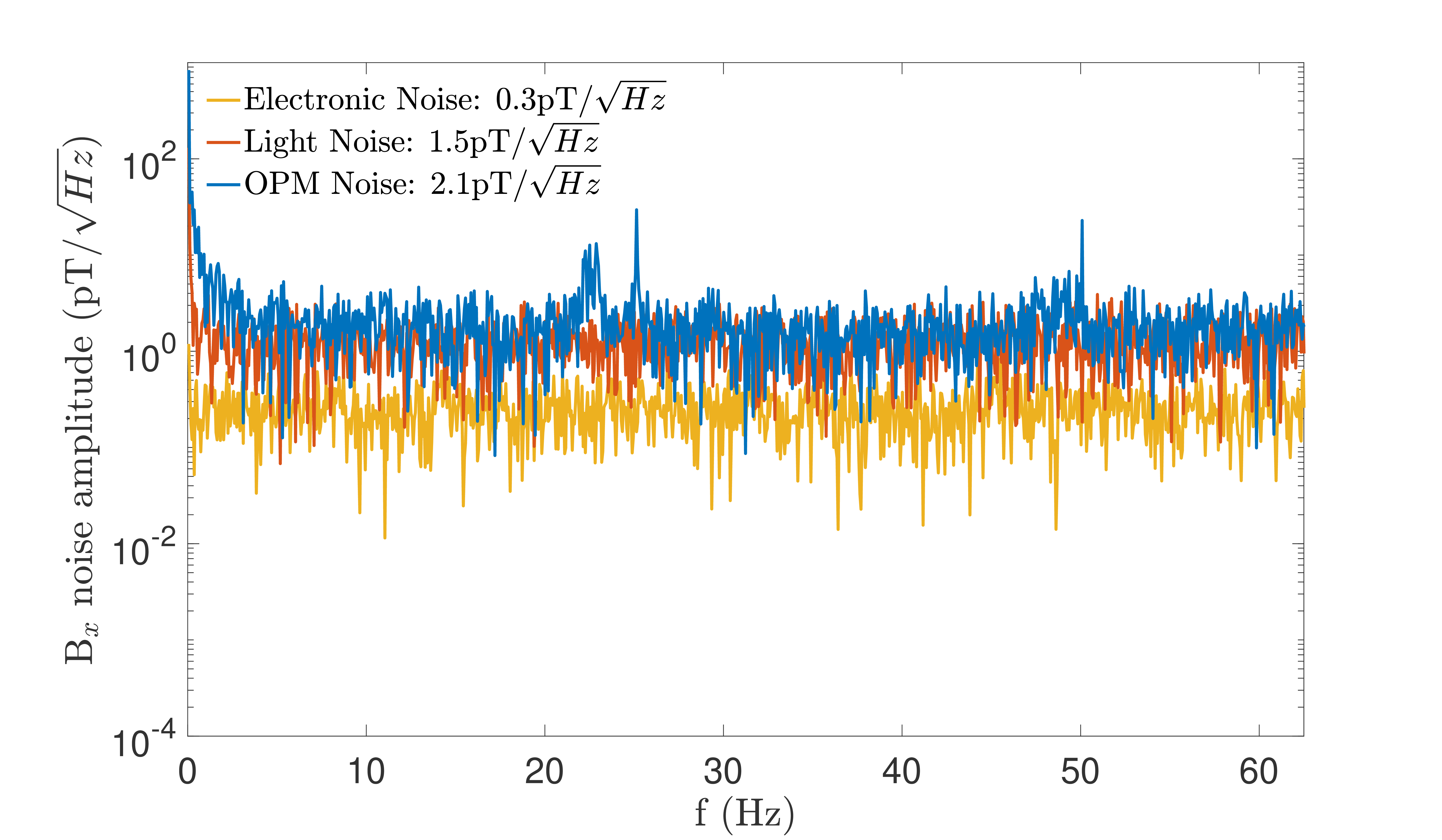}
\put(3,53){a)}
\end{overpic}
\hspace*{-0.3cm}
\begin{overpic}[width=\textwidth/2]{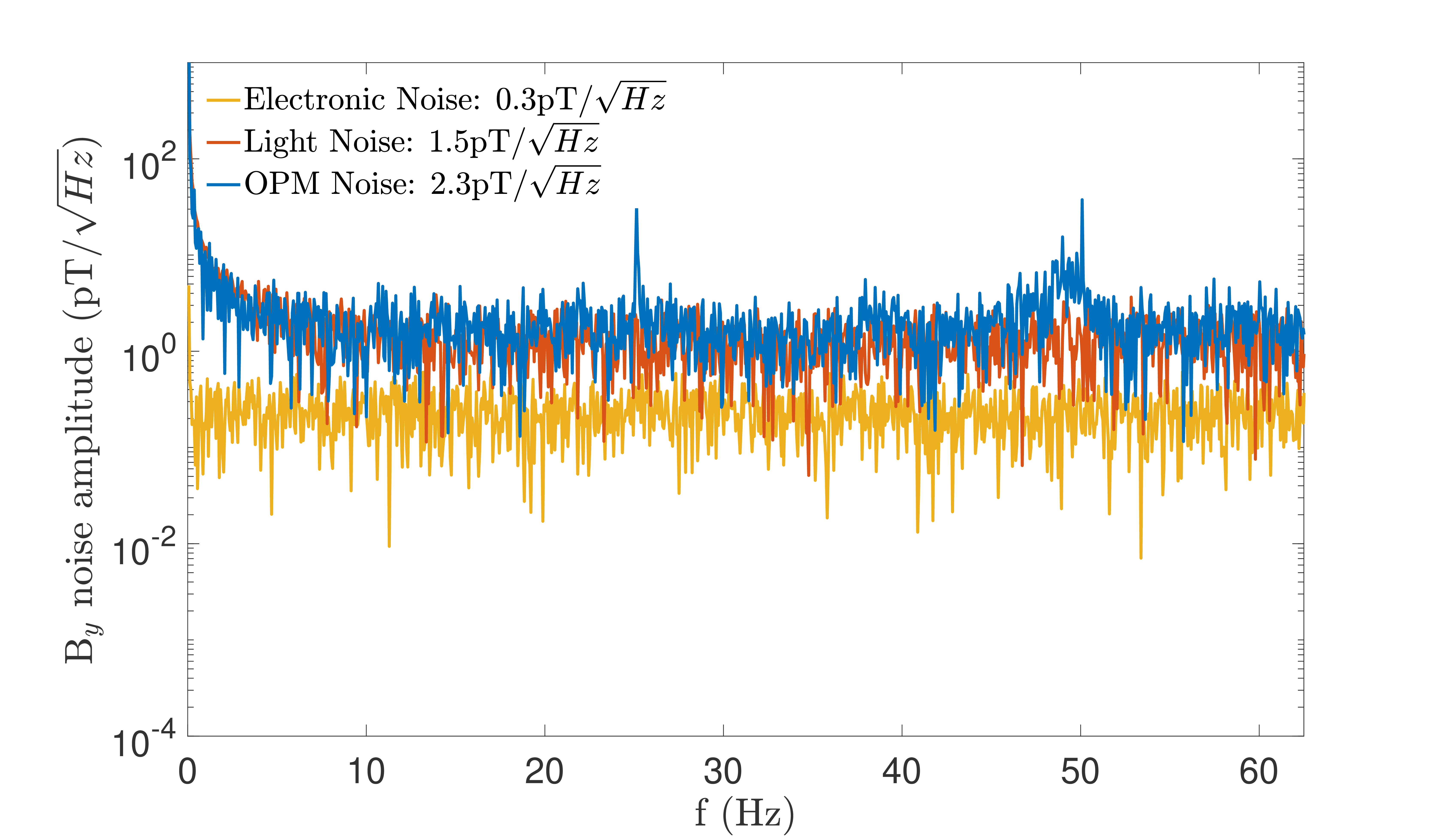}
\put(3,53){b)}
\end{overpic}
\hspace*{-0.3cm}
\begin{overpic}[width=\textwidth/2]{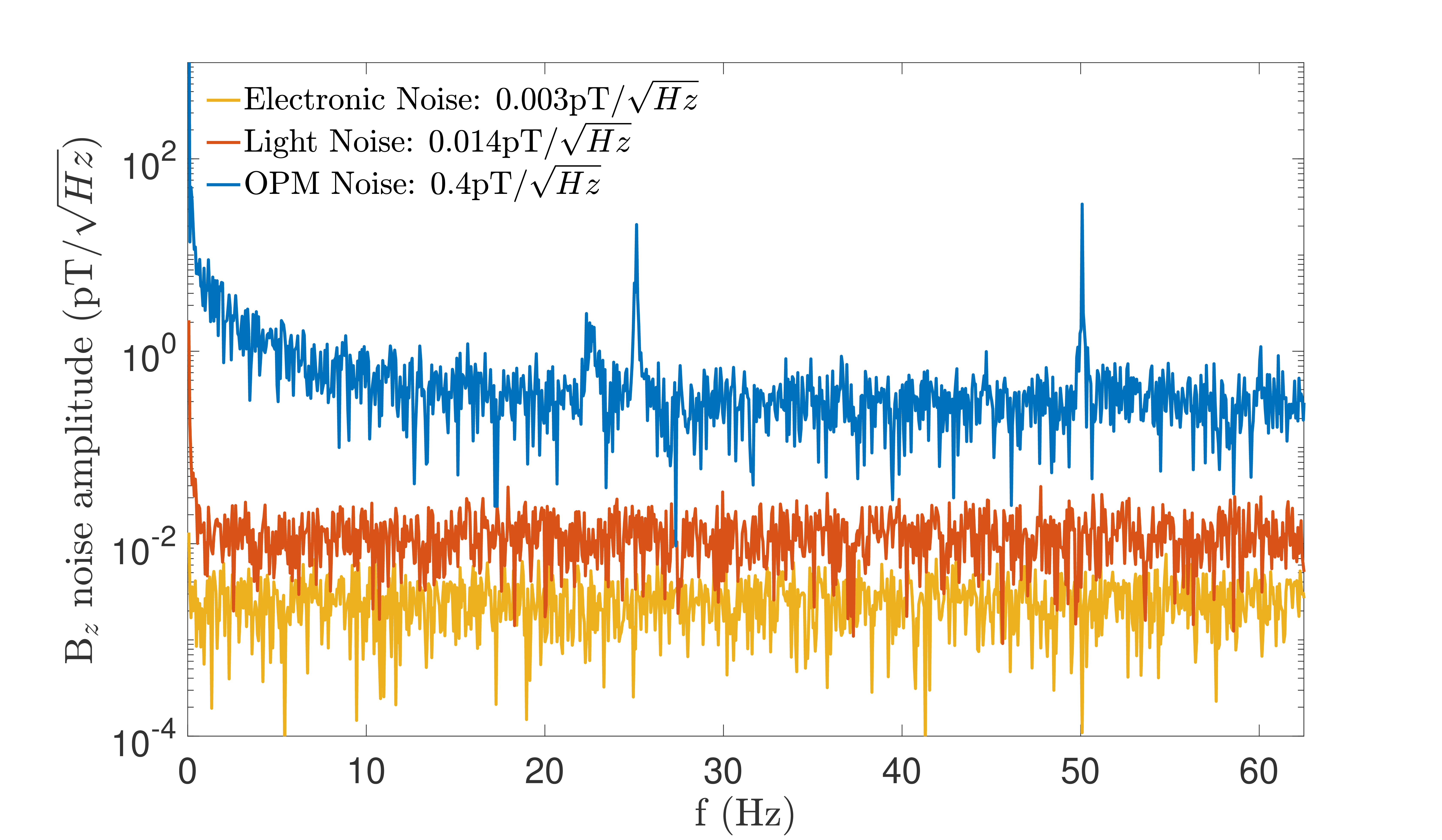}
\put(3,53){c)}
\end{overpic}
\caption{OPM noise at 36$^{\circ}$C vapour cell temperature in a shielded environment at 5 kHz radio frequency dressing. Noise floor values are estimated for the range 10~Hz~-~62.5~Hz. Panels (a) and (b) show noise performance for the two orthogonal transverse fields. The light noise in one of the signal quadratures shows phase-locked, low-frequency fluctuations of cross talk between the rf generation and detection paths. Panel (c) shows the noise performance for the longitudinal field component. The light noise levels (photon shot noise) are obtained with a far-detuned probe laser and disabled pump/repump lasers. The electronic noise is recorded without probe light and no rf field present.
The calibration of field equivalent noise amplitudes includes an $\approx5\%$-drop of the low-pass frequency response function, which is predominantly determined by the mode function entering Eq.~(\ref{eq:mode_amplitudes}).} 
\label{fig:opm_noise}
\end{figure} 
\begin{figure}[h]\centering\hspace*{-0.5cm}
\begin{overpic}[width=\textwidth/2]{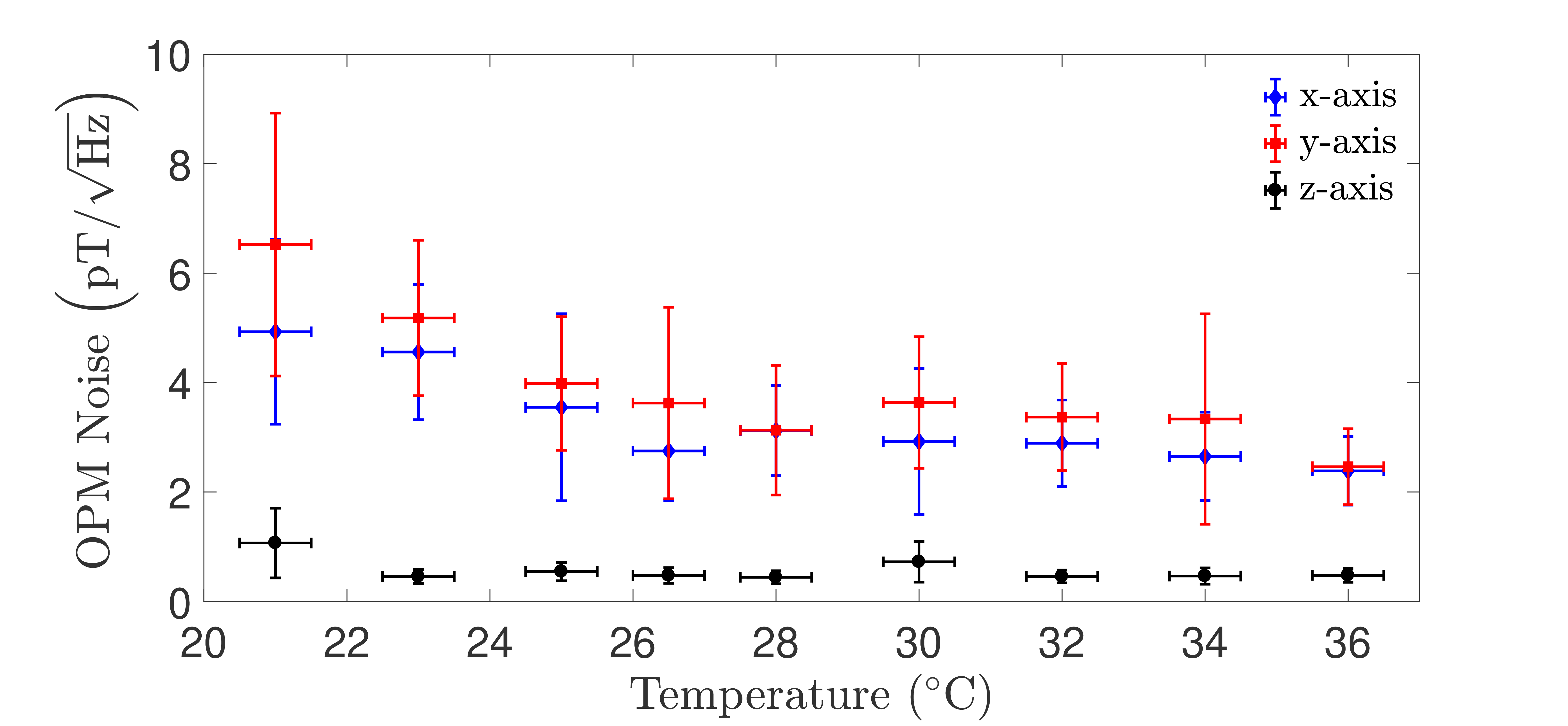}
\put(5,43){a)}
\end{overpic}
\hspace*{-0.5cm}
\begin{overpic}[width=\textwidth/2]{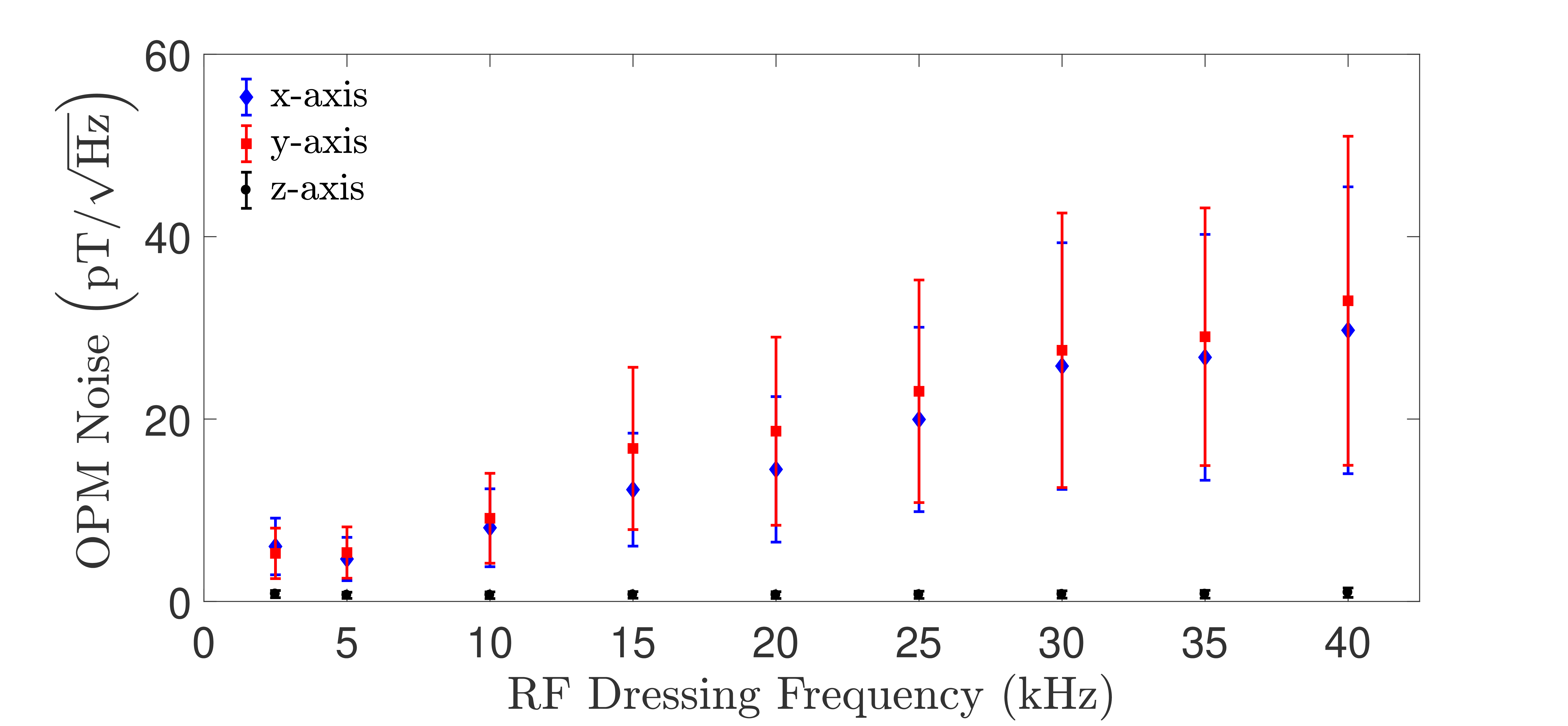}
\put(5,43){b)}
\end{overpic}
\caption{Dependencies of the noise performance. (a) OPM noise as a function of the vapour cell temperature. (b) OPM noise as a function of radio frequency of the dressing field.} \label{fig:opm_noise_vs_RF}
\end{figure}
Based on the field calibrations described above, we record field equivalent signal noise for the three field components over $\approx16$~s (2048 cycles at 125~Hz).
Figure~\ref{fig:opm_noise} shows the spectral noise performance
for the two quadratures at $\omega$ and the in-phase quadrature at $2\omega$.
At 5 kHz rf dressing frequency and a temperature of $36^\circ$C,
the magnetometer operates with an average noise level of $\approx$ 2.2~pT$/\sqrt{\mathrm{{Hz}}}$ for the transverse fields over the range of 10-62.5~Hz, dominated by photon shot noise. Longitudinal fields can be measured with a sensitivity of 0.4~pT$/\sqrt{\mathrm{{Hz}}}$. The dominant constraint on the noise level is the short coherence time of the cell ($\tau\approx 2$~ms) which is limited by the quality of the paraffin coating and the exchange of the atoms between the main cell  body  and  the  stem  with  the  Rb  reservoir. Typically, paraffin or OTS coated cells have coherence times ranging from 30 ms to 300 ms~\cite{seltzer,coating}. Longer coherence time would improve the field sensitivity of the OPM due to a larger fraction of atoms remaining in the field sensitive state. In addition, higher quality paraffin coating would also shorten the pump/repump pulse time needed to (re-)prepare the stretched states, allowing for increased cycle rate and thus higher bandwidth as well as higher duty cycle and thus reduced aliasing of magnetic field noise. 

We have investigated the effects of heating the cell to increase the atomic density, which should in principle improve the sensitivity by increasing signal strength. However, as it is shown in Fig.~\ref{fig:opm_noise_vs_RF}(a), the signal-to-noise-ratio saturates already at temperatures of approximately $32^\circ$C for the transverse fields and at even lower temperatures for the longitudinal fields. Initially, increasing signal amplitudes lead to better sensitivity, especially for the transverse fields where the signals are closer to photon shot noise. But additional atomic processes such as resonance broadening limit the performance at higher temperature where the figure of merit saturates. A similar saturation effect was previously  observed in ref.~\cite{pustelny}, where higher atomic concentrations lead to an increase of the collisional and surface relaxation rates, depolarizing the prepared state.

The sensitivity of the OPM to transverse fields does not only depend on the shape of the resonance, but also on the chosen dressing frequency, because the corresponding signals arise from the geometric rotation of the static field. The rotation angle and consequently the signal strength increases for smaller offset fields, see Eqs.~(\ref{Re1f}) and (\ref{Im1f}). The resulting linear dependence of sensitivity on dressing frequency is shown in Fig.~\ref{fig:opm_noise_vs_RF}(b). Over the range of 40~kHz to 2.5~kHz the transverse field noise performance varies by a factor of four. This strategy is limited by the linewidth of the rf resonance and other factors such as the required precision of alignment increased susceptibility to magnetic field gradients distorting the oviform mapping. 

\section{\label{sec:Conclusions}Conclusions}

We have presented and successfully demonstrated a full vector magnetometer based on the Voigt effect both in cold atom and hot vapour setups. As shown, our scheme has the advantage of requiring only a single optical axis for state preparation and detection making it ideal for compact magnetic field sensors. We have achieved pT/$\sqrt{\mathrm{Hz}}$ sensitivity with a $62.5$ Hz bandwidth. Our current limitations in the sensitivity of the OPM stem from the coherence time of the cell and the low atom number. Future improvements will include a heated and buffer gas filled cell in order to increase the atom numbers and reduce the rate of atomic collisions that induce decoherence effects, respectively. These improvements should shorten the state preparation lifetime thus increasing the bandwidth and the field sensitivity of the OPM. In principle, placing the cell in an optical resonator may be used to increase the interaction path between the light and the atoms thus further improving the sensitivity. 

The datasets generated for this paper are accessible at \cite{data} Nottingham Research Data Management Repository.

\section{Acknowledgements}

This work was funded by Engineering and Physical Sciences
Research Council (Grant No. EP/M013294/1). We acknowledge the support from the School of Physics \& Astronomy Engineering and Electronics workshops. We thank Sindhu Jammi for collecting the cold atoms data. We thank Konstantinos Poulios and Kasper Jensen for useful discussions.

\appendix

\section{\label{sec:faraday}Faraday and Voigt rotation for dressed states}

In what follows we present a brief description
of the atom dynamics
and the resulting behaviour of Faraday and Voigt rotation,
following ref.~\cite{Jammi}.

Starting from our effective, rotating frame Hamiltonian in Eq.~(\ref{eq:Heff}), the dressed states are of the form 
\begin{equation}
    \left|\Psi_{\mathrm{rot}}\right\rangle=e^{i\theta\hat{F}_y/\hbar}\left|F,F_z\right\rangle,
\end{equation}
where $\theta$ is the angle of the effective magnetic field direction, given in Eq.~(\ref{eq:theta}). In the laboratory frame,
the same states are given by 
\begin{equation}
\left|\Psi(t)\right\rangle=\hat{U}_{\pm}^{-1}(t)\left|\Psi_{\mathrm{rot}}\right\rangle,
\end{equation}
where $\hat{U}_{\pm}(t)=e^{\pm i\omega t \hat{F_z}/\hbar}$. Combining the two rotation operators, the expectation value of any laboratory frame, atomic observable can be expressed as
\begin{align}
    \left\langle\Psi(t)\right|\hat{O}\left|\Psi(t)\right\rangle&=\left\langle F,F_z\right|\hat{R}_{\pm}\hat{O}\hat{R}_{\pm}^{-1}\left|F,F_z\right\rangle,\label{eq:Mv_Obs}\\
    \hat{R}_{\pm}(t)&=e^{-i\theta\hat{F}_y/\hbar}\hat{U}_{\pm}(t). 
\end{align}
Using the Baker--Hausdorff Lemma
\begin{equation}
e^{\hat{A}}\hat{B}e^{-\hat{A}}=\hat{B}+[\hat{A},\hat{B}]+\frac{1}{2!}[\hat{A},[\hat{A},\hat{B}]]+...,
\end{equation}
it can be shown that for a Cartesian vector operator, the corresponding Heisenberg operator is given by a geometric rotation, i.e.\ 
\begin{equation}
\hat{\mathbf{O}}'(t)=\hat{R}_{\pm}(t)\hat{\mathbf{O}}\hat{R}_{\pm}^{-1}(t)=\mathbf{{R}_{\pm}}(t)\hat{\mathbf{O}},
\end{equation}
where $\mathbf{{R}_{\pm}}(t)=\mathbf{{R}_{z}}(\pm\omega t)\mathbf{{R}_{y}}(-\theta)$ \cite{Rodrigues}. Including the rotations of the static field as shown in Fig.~\ref{fig:FrameRotation}, such that $\hat{\mathbf{F}}'(t)=\mathbf{M}(\alpha,\beta)\hat{\mathbf{R}}_\pm(t)\hat{\mathbf{F}}$,
we find an expression for the longitudinal spin component
\begin{align}
\Braket{\hat{F}'_z(t)}=&\Braket{\hat{F}_x}\left(c_\alpha c_\beta s_\theta-c_\theta (c_{\omega t} s_\beta+c_\beta s_\alpha s_{\omega t} ) \right)\nonumber\\
&+\Braket{\hat{F}_y}\left(c_\beta c_{\omega t} s_{\alpha}-s_{\beta} s_{\omega t}  \right)\nonumber\\
&+\Braket{\hat{F}_z}\left(c_\alpha c_\beta c_\theta+s_\theta (c_{\omega t} s_\beta+c_\beta s_\alpha s_{\omega t}) \right),\label{eq:Fz}
\end{align}
where $s_{\nu}$ ($c_{\nu}$) stands for $\sin(\nu)$ ($\cos(\nu)$) with $\nu~\in~\{\alpha,\beta, \omega t \}$. For the dressed states, i.e.\ for eigenstates of the rotating frame Hamiltonian, the relevant expectation values $\Braket{\hat{F}_{x,y,z}}=\langle F,F_z|\hat{F}_{x,y,z}|F,F_z\rangle$ are constant, with $\Braket{\hat{F}_z}=\hbar F_z$ and $\Braket{\hat{F}_{x,y}}=0$.
The longitudinal spin polarization determines the Faraday rotation, see Eq.~(\ref{eq:Faraday}), and we obtain
\begin{equation}
\Braket{\hat{S}_x'(t)}
=-G_{F}^{(1)} S_y n_F \hbar F_z \left(c_\alpha c_\beta c_\theta+s_\theta (c_{\omega t} s_\beta+c_\beta s_\alpha s_{\omega t})\right).
\end{equation}
The result can be expanded in terms of spectral components as
\begin{equation}
\Braket{\hat{S'}_x(t)}=-\frac{1}{2}G_{F}^{(1)}S_y n_F\hbar F_z \sum_{n=0}^1 \tilde{h}_n(\theta)e^{in\omega t}+c.c.,
\end{equation}
where the spectral amplitudes vary as
\begin{align}
\tilde{h}_0&=\cos\alpha\cos\beta\cos\theta, \label{eq:h0Fz}\\
\tilde{h}_1&=(\sin\beta\pm i\sin\alpha\cos\beta)\sin\theta. \label{eq:h1Fz}
\end{align}
Equivalently, for the Voigt Rotation, in which the elliptical light polarization is given by Eq.~(\ref{eq:Voigt}), the spectral decomposition of the mean value in Eq.~(\ref{eq:Mv_Obs}) for the bi-linear operator $\hat{\mathbf{O}}^{\dagger}\hat{\mathbf{O}}=\hat{F}_x^2-\hat{F}_y^2$ is 
\begin{equation}
\Braket{\hat{S'}_z(t)}=\frac{1}{2}G_{F}^{(2)}S_y n_F\hbar^2\xi_F(F_z)\sum_{n=0}^2 h_n(\theta)e^{in\omega t}+c.c.,
\label{eqn:spectraloutput}
\end{equation}
where the spectral amplitudes vary as
\begin{align}
h_0&=\frac{1+3\cos{2\theta}}{4}
        \left(\frac{\cos^2\!{\beta}}{2}{\scriptstyle -}\frac{(3-\cos{2\beta})}{4}\cos{2\alpha}
        \right), \label{eq:h0Fxy}\\
h_1&=\bigg(\frac{1}{2}\cos\alpha\sin 2\beta \mp i\frac{(3-\cos 2\beta)}{4}\sin 2\alpha\bigg)\sin2\theta,\label{eq:h1Fxy} \\ 
h_2&=-\bigg(\frac{(3-\cos 2\beta)}{4}\cos^2\alpha+\frac{1}{2}\cos 2\beta \notag \\ 
    &\hspace{3.2cm} \mp\frac{i}{2}\sin\alpha\sin 2\beta \bigg) \sin^2\theta.
\label{eq:h2Fxy}
\end{align}

In order to measure an external magnetic field, we extract three quadratures from the first and second harmonic signals. We separate the first harmonic into real and imaginary parts, and evaluate the real part of the second harmonic. The result is given by
\begin{align}
h_x=&\Re{h_1}=\frac{1}{4}(3-\cos 2\beta)\sin 2\alpha\sin 2\theta, \\
h_y=&\Im{h_1}=\frac{1}{2}\cos\alpha\sin 2\beta\sin 2\theta, \\
h_z=&\Re{h_2}=-\frac{1}{4}\left( (3-\cos 2\beta)\cos^2\alpha +2\cos 2\beta \right)\sin^2\theta.
\end{align}

%

\begin{thebibliography}{41}
\bibitem{Rodrigues}
We use rotation matrices according to Rodrigues's rotation formula $\mathbf{R}_k(\alpha)\mathbf{v}=\mathbf{e}_k(\mathbf{e}_k\cdot\mathbf{v})(1-\mathrm{cos}\alpha)+\mathbf{v}\mathrm{cos}\alpha+\mathbf{e}_k\cross\mathbf{v}\mathrm{sin}\alpha$.
\end{thebibliography}
\end{document}